\documentclass[acmsmall,screen]{acmart}
\usepackage{epsfig,endnotes}
\usepackage{xspace}
\usepackage{subcaption}
\usepackage{hyperref}
\usepackage{color}
\usepackage{tabularx}
\usepackage{booktabs}
\usepackage{graphicx}
\usepackage{amsmath}
\usepackage{pifont}
\usepackage{wrapfig}

\setcopyright{acmlicensed}
\acmPrice{}
\acmDOI{10.1145/3276518}
\acmYear{2018}
\copyrightyear{2018}
\acmJournal{PACMPL}
\acmVolume{2}
\acmNumber{OOPSLA}
\acmArticle{148}
\acmMonth{11}

\usepackage{booktabs}
\usepackage{subcaption}
\usepackage{clrscode3e}
\usepackage{xspace}
\usepackage{microtype}
\usepackage[super]{nth}
\usepackage{mathtools}
\usepackage{amsfonts}
\usepackage{amsmath}
\usepackage{amsthm}
\usepackage{stmaryrd}
\usepackage{multirow}
\usepackage{hhline}
\usepackage{todonotes}
\usepackage[font=small,labelfont=bf,skip=0pt]{subcaption}
\usepackage{listings}
\usepackage{microtype}
\usepackage{placeins}
\usepackage{afterpage}
\usepackage{url}
\usepackage{balance}

\SetTracking{encoding={*}, shape=sc}{0}
\lstdefinestyle{Common}
{
    extendedchars=\true,
    frame=single,
    framesep=3pt,
    framerule=0.4pt,
    xleftmargin=3.4pt,
    xrightmargin=3.4pt,
    basicstyle=\ttfamily\footnotesize,
    aboveskip=0pt,
    belowskip=5pt
}

\newcommand{\sectref}[1]{\S\ref{#1}}

\newcommand{\flashrelate}{\textsc{FlashRelate}\xspace}

\newcommand{\excelint}{\textsc{ExceLint}\xspace}

\newcommand{\custodes}{\textsc{CUSTODES}\xspace}

\newcommand{\idthree}{\textsc{ID3}\xspace}
\newcommand{\faultysheet}{\textsc{FaultySheet Detective}\xspace}

\DeclareMathOperator*{\argmin}{\arg\!\min}

\newcommand{\AnnoHours}{34\xspace}
\newcommand{\numexsheets}{70\xspace}
\newcommand{\exmedianruntime}{5\xspace}
\newcommand{\ExMedianRuntimeSeconds}{\exmedianruntime}
\newcommand{\NumSmells}{1,199\xspace}
\newcommand{\CUSTODESNumTrueRefBugs}{102\xspace}
\newcommand{\ExceLintNumTrueRefBugs}{397\xspace}
\newcommand{\ECTrueRefBugDelta}{295\xspace}
\newcommand{\TotalAnno}{9,924\xspace}
\newcommand{\ExceLintMinutesPerBug}{5.1\xspace}

\newcommand{\ExceLintMeanPrecisionPercent}{64.1\%\xspace}

\newcommand{\ExceLintMeanAdjustedPrecisionPercent}{63.7\%\xspace}
\newcommand{\ExceLintMedianPrecisionPercent}{100\%\xspace}
\newcommand{\ExceLintMedianRecallPercent}{100\%\xspace}

\newcommand{\ExceLintMeanRecallPercent}{62.1\%\xspace}

\newcommand{\ExceLintRandomNegativeCases}{7\xspace}

\newcommand{\ExceLintTotalTP}{89\xspace}
\newcommand{\ExceLintTotalFP}{223\xspace}

\newcommand{\CUSTODESMeanPrecisionPercent}{20.3\%\xspace}

\newcommand{\CUSTODESMeanAdjustedPrecisionPercent}{19.6\%\xspace}
\newcommand{\CUSTODESMedianPrecisionPercent}{0.0\%\xspace}
\newcommand{\CUSTODESMedianRecallPercent}{100.0\%\xspace}

\newcommand{\CUSTODESMeanRecallPercent}{61.2\%\xspace}

\newcommand{\CUSTODESRandomNegativeCases}{14\xspace}

\newcommand{\CUSTODESTotalTP}{52\xspace}
\newcommand{\CUSTODESTotalFP}{1,816\xspace}
\newcommand{\WastedEffort}{$8\times$\xspace}
\newcommand{\ExceLintPrecisionPointDelta}{44.9\xspace}
\newcommand{\ExcelintNumDoNothing}{33\xspace}
\newcommand{\ExcelintNumDoNothingCorrectly}{22\xspace}
\newcommand{\NumHighFPBenchmarks}{5\xspace}
\newcommand{\PctHighFPBenchmarksExcelint}{45.2\%\xspace}
\newcommand{\PctHighFPBenchmarksCUSTODES}{64.3\%\xspace}
\newcommand{\PctSSFormulaMeanRectangular}{86.8\%\xspace}
\newcommand{\PctSSFormulaMedianRectangular}{90.5\%\xspace}
\newcommand{\PctSSMeanRectangular}{62.3\%\xspace}
\newcommand{\PctSSMedianRectangular}{63.2\%\xspace}

\newcommand{\ExceLintMedianRuntime}{\exmedianruntime}
\newcommand{\ExceLintMeanRuntime}{14.0\xspace}

\newcommand{\ExceLintPctUnderThirtySec}{85.2\%\xspace}

\newcommand{\meanHashCollisionsPct}{0.33\%\xspace}
\newcommand{\medianHashCollisionsPct}{0.0\%\xspace}

\newcommand{\SuiteMinCells}{99\xspace}
\newcommand{\SuiteMaxCells}{12,121\xspace}
\newcommand{\SuiteMeanCells}{1580\xspace}
\newcommand{\SuiteMinFormulas}{3\xspace}
\newcommand{\SuiteMaxFormulas}{2,662\xspace}
\newcommand{\SuiteMeanFormulas}{360\xspace}

\newcommand{\punt}[1]{}

\begin{document}

\title[\excelint: Automatically Finding Spreadsheet Formula Errors]{\excelint: Automatically Finding Spreadsheet Formula Errors}


\author{Daniel W. Barowy}
\affiliation{
  \department{Department of Computer Science}              
  \institution{Williams College}            
  \country{USA}                    
}
\email{dbarowy@cs.williams.edu}          

\author{Emery D. Berger}
\affiliation{
  \department{College of Information and Computer Sciences}              
  \institution{University of Massachusetts Amherst}            
  \country{USA}                    
}
\email{emery@cs.umass.edu}          

\author{Benjamin Zorn}
\affiliation{
  \institution{Microsoft Research}            
  \country{USA}                    
}
\email{ben.zorn@microsoft.com}          

\begin{abstract}

\noindent
Spreadsheets are one of the most widely used programming environments,
and are widely deployed in domains like finance where errors can have
catastrophic consequences. We present a static analysis specifically
designed to find spreadsheet formula errors. Our analysis directly
leverages the rectangular character of spreadsheets. It uses an
information-theoretic approach to identify formulas that are
especially surprising disruptions to nearby rectangular regions.  We
present \excelint{}, an implementation of our static analysis for Microsoft Excel.
We demonstrate that \excelint{} is fast and effective:
across a corpus of 70 spreadsheets, \excelint{}
takes a median of \ExceLintMedianRuntime seconds per spreadsheet, and it significantly outperforms the state of the art analysis.

\end{abstract}



\maketitle

\section{Introduction}
\label{sec:intro}
In the nearly forty years since the release of VisiCalc in 1979,
spreadsheets have become the single most popular end-user programming
environment, with 750 million users of Microsoft Excel
alone~\cite{750mil}.  Spreadsheets are ubiquitous in government,
scientific, and financial settings~\cite{Panko98whatwe}.

Unfortunately, errors are alarmingly common in spreadsheets: a 2015
study found that more than 95\% of spreadsheets contain at least one
error~\cite{PankoEUSPRIG2015}. Because spreadsheets are frequently
used in critical settings, these errors have had serious
consequences. For example, the infamous ``London Whale'' incident in
2012 led J.P. Morgan Chase to lose approximately \$2 billion (USD) due
in part to a spreadsheet programming error~\cite{whale}. A Harvard
economic analysis used to support austerity measures imposed on
Greece after the 2008 worldwide financial crisis was based on a
single large spreadsheet~\cite{NBERw15639}. This analysis was later
found to contain numerous errors; when fixed, its conclusions were
reversed~\cite{herndon2013does}.


Spreadsheet errors are common because they are both easy to introduce
and difficult to find. For example, spreadsheet user interfaces make it simple for
users to copy and paste formulas or to drag on a cell to fill a
column, but these can lead to serious errors if references are not
correctly updated. Manual auditing of formulas is time consuming and
does not scale to large sheets.




\subsection{Contributions}

Our primary motivation behind this work is to develop static analyses, based on principled statistical techniques, that automatically find errors in spreadsheets without user assistance and with high median precision and recall.  This paper makes the following contributions.

\begin{itemize}
	\item \excelint's analysis is the first of its kind, operating without annotations or user guidance; it relies on a novel and principled information-theoretic static analysis that obviates the need for heuristic approaches like the bug pattern databases used by past work.  Instead, it identifies formulas that cause surprising disruptions in the distribution of rectangular regions.  As we demonstrate, such disruptions are likely to be errors.
	\item  We implement \excelint for Microsoft Excel and present an extensive evaluation using a commonly-used representative corpus of \numexsheets benchmarks (not assembled by us) in addition to a case study against a professionally audited spreadsheet. When evaluated on its effectiveness at finding real formula errors, \excelint outperforms the state of the art, \custodes, by a large margin.  \excelint is fast (median seconds per spreadsheet: \ExceLintMedianRuntime), precise (median precision: \ExceLintMedianPrecisionPercent), and has high recall (median: \ExceLintMedianRecallPercent).
\end{itemize}

%
%
%


\begin{figure*}[!t]
\centering
\begin{subfigure}{0.5\linewidth}
  \centering
  \includegraphics[width=0.97\linewidth]{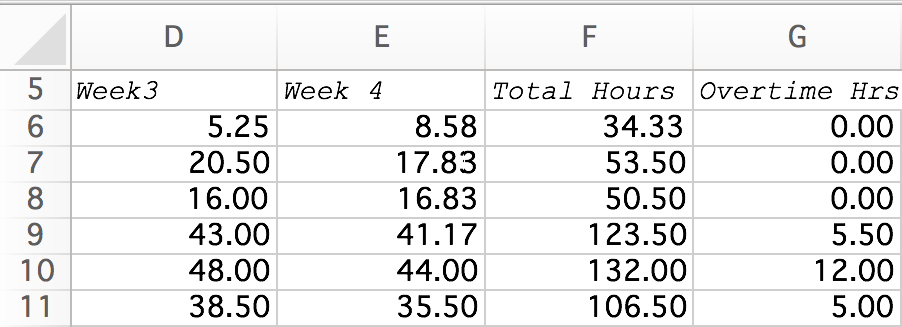}
  \caption{}
  \label{fig:Act3}
\end{subfigure}%
\begin{subfigure}{0.5\linewidth}
\centering
  \includegraphics[width=0.75\linewidth]{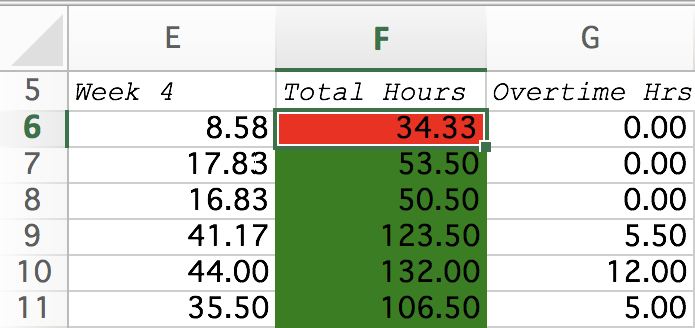}
  \caption{}
  \label{fig:Act3ProposedFix}
\end{subfigure}	
\caption{\textbf{\excelint{} in action.} An excerpt of a buggy spreadsheet drawn from the CUSTODES corpus~\cite{custodes, Fisher:2005:ESC:1082983.1083242}.  (a) In Excel, errors are not readily apparent. (b) Output from \excelint{}: a suspected error is shown in red, and the proposed fix is shown in green. This is an actual error: the formula in \texttt{F6}, \texttt{=SUM(B6:E6)}, is inconsistent with the formulas in \texttt{F7:F11}, which omit \texttt{Week 4}.}
\label{fig:test}
\end{figure*}

\section{Overview}
\label{sec:overview}
This section describes at a high level how \excelint's static analysis works.

Spreadsheets strongly encourage a rectangular organization scheme.
Excel's syntax makes it especially simple to use rectangular
regions via so-called \emph{range references}; these refer to
groups of cells (e.g., \texttt{A1:A10}).  Excel also comes with a
large set of built-in functions that make operations on ranges
convenient (e.g., \texttt{SUM(A1:A10)}).  Excel's user interface,
which is tabular, also makes selecting, copying, pasting, and
otherwise manipulating data and formulas easy, as long as related
cells are arranged in a rectangular fashion.

The organizational scheme of data and operations on a given worksheet
is known informally as a \emph{layout}.  A \emph{rectangular layout}
is one in which related data or related operations are placed adjacent
to each other in a rectangular fashion, frequently in a column. Prior
work has shown that users who eschew rectangular layouts find
themselves unable to perform even rudimentary data analysis
tasks~\cite{Barowy:2015:FER:2737924.2737952}. Consequently,
spreadsheets that contain formulas are almost invariably rectangular.

\excelint exploits the intrinsically
rectangular layout of spreadsheets to identify formula errors.  The analysis first
constructs a model representing the rectangular layout intended by the
user.  

\begin{wrapfigure}{r}{0.5\textwidth}
  \begin{centering}
    \includegraphics[width=\linewidth]{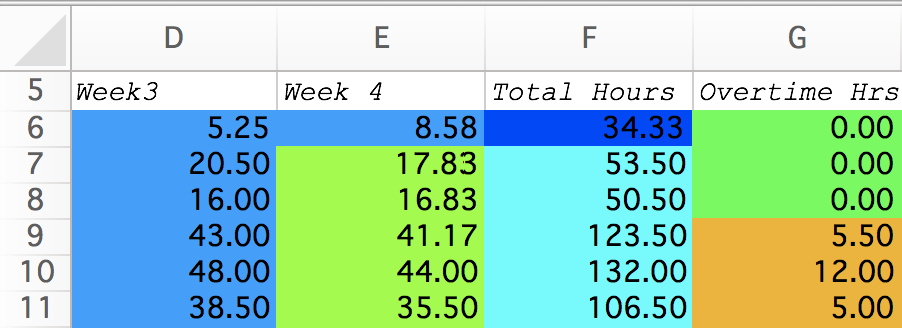}
  \end{centering}
  \caption{Fingerprint regions for the same spreadsheet shown in Figure~\ref{fig:Act3}.  Note that we color numeric data here the same shade of blue (column \texttt{D} and \texttt{E6}) to simplify the diagram. See \sectref{sec:GlobalView} for details.\label{fig:regions}}  
\end{wrapfigure}

Since there are many possible layouts and because user intent
is impossible to know, \excelint uses simplicity as a proxy: the
simplest layout that fits the data is most likely the intended layout.
In this setting, formula errors manifest as
aberrations in the rectangular layout.  To determine whether such an
aberration is likely to be an error, \excelint uses the cell's
position in the layout (that is, its context) to propose a ``fix'' to
the error.  If the proposed fix makes the layout
simpler---specifically, by minimizing the entropy of the distribution
of rectangular regions---then the cell is flagged as a suspected
error.

The remainder of this section provides an overview of how each phase
of \excelint's analysis proceeds.

\subsection{Reference Vectors}
\excelint compares formulas by their \emph{shape} rather than
syntactically; mere syntactic differences are often insufficient to
distinguish the different computations. Consider a column of formulas
in column \texttt{C} that all add numbers found in the same row of
columns \texttt{A} and \texttt{B}. These formulas might
be \texttt{=A1+B1}, \texttt{=A2+B2}, \texttt{=SUM(A3:B3)}, and so
on. Each is syntactically different but semantically identical.

An important criterion used in \excelint's analysis is the similarity of adjacent formulas.  Nearly similar formulas often indicate an error. Large differences between formulas usually indicate entirely different computations.  As a result, \excelint needs a way to measure the ``distance'' between formulas, like \texttt{=A1+B1} and \texttt{=A1}.


\excelint{} uses a novel
vector-based representation of formulas we call \emph{reference
vectors} that enable both reference shape and formula distance comparisons. Reference vectors achieve this by unifying spatial and dependence information into a single geometric construct relative to the given formula (\sectref{sec:vectorcompute}). Consequently, formulas that exhibit similar reference behaviors induce the same set of reference vectors.

Consider the formula \texttt{=A1+B1} located in
cell \texttt{C1}: it has two referents, the cells \texttt{A1}
and \texttt{B1}, so the reference vector for this formula consists of
two vectors, \texttt{C1}$\rightarrow$\texttt{A1}
and \texttt{C1}$\rightarrow$\texttt{B1}. \excelint{} transforms these
references into offsets in the Cartesian plane with the origin at the
top left, $(-2, 0)$ and $(-1, 0)$. As a performance optimization, the
analysis compresses each formula's set of vectors into a resultant
vector that sums its component vectors; resultants are used instead of vector sets for formula comparisons.  We call this compressed representation a \emph{fingerprint} (\sectref{sec:fingerprintoptim}).

\excelint{} extracts reference vectors by first gathering data
dependencies for every formula in the given spreadsheet. It obtains
dependence information by parsing a sheet's formulas and building the
program's dataflow
graph~\cite{torczon,Barowy:2014:CDD:2660193.2660207}.  \excelint can analyze all Excel functions.


\subsection{Fingerprint Regions}

As noted above, the syntax and user interfaces of spreadsheets
strongly encourage users to organize their data into rectangular
shapes.  As a result, formulas that observe the same relative spatial
dependence behavior are often placed in the same row or column.
The second step of \excelint{}'s analysis identifies homogeneous,
rectangular regions; we call these \emph{fingerprint regions} (\sectref{sec:decomp} and \sectref{sec:fingerprintoptim}). These
regions contain formulas with identical fingerprints, a proxy for
identical reference behavior.  Figure~\ref{fig:regions} shows a set of fingerprint regions for the spreadsheet shown in Figure~\ref{fig:Act3}.

\excelint computes fingerprint regions via a top-down,
recursive decomposition of the spreadsheet.  At each step, the
algorithm finds the best rectangular split, either horizontally or vertically.
This procedure is directly inspired by the \idthree decision tree
algorithm~\cite{Quinlan86inductionof}.  The algorithm greedily partitions a space into a collection of rectangular regions. Once this
decomposition is complete, the result is a set of regions guaranteed
to be rectangular, homogeneous (consisting of cells with the
same fingerprint), and be a low (near optimal) entropy decomposition of the plane.


\subsection{Candidate Fixes and Fix Ranking}
\excelint{} identifies \emph{candidate fixes} by
comparing cells to all adjacent rectangular regions
(\sectref{sec:ProposedFix}). Each candidate fix is a pair composed of one or more suspect formulas and a neighboring set of formulas that exhibit different reference behavior.  We call the pair a ``fix'' because it suggests a way to update suspect formulas such that their fingerprints match the fingerprints of their neighbors.  The outcome of applying the fix is the creation of a larger rectangular
region of formulas that all exhibit the same reference behavior.

The highest ranked candidate fixes pinpoint those formulas that are both similar to their neighbors and cause small drops in entropy when ``fixed.''  Intuitively, such differences are likely to be the product of an error like failing to update a reference after pasting a
formula. Because differences are small, they are easy to miss during spot checks. Large
differences between pairs are usually not indicative of error; more often, they are simply neighboring regions that deliberately perform a different calculations.

\subsection{Errors and Likely Fixes}
Finally, after ranking, \excelint{}'s user interface guides users
through a cell-by-cell audit of the spreadsheet, starting with the
top-ranked cell (\sectref{sec:visualizations}).  Because broken
formula behaviors are difficult to understand out of
context, \excelint{} visually pairs errors with their likely proposed fixes, as
shown in Figure~\ref{fig:Act3ProposedFix}.

\section{\excelint{} Static Analysis}
\label{sec:algorithms}
This section describes
\excelint{}'s static analysis algorithms in detail.

\subsection{Definitions}
\label{sec:defs}

\paragraph{Reference vectors:}
a \emph{reference vector} is the basic unit of analysis in \excelint.
It encodes not just the data dependence between two cells in a
spreadsheet, but also captures the spatial location of each def-use
pair on the spreadsheet.  Intuitively, a reference vector can be
thought of as a set of arrows that points from a formula to each of the
formula's inputs.  Reference vectors let \excelint's analysis determine whether two
formulas point to the same relative offsets.  In essence, two formulas
are \emph{reference-equivalent} if they induce the same vector set.

Reference vectors abstract over both the operation utilizing the
vector as well as the effect of copying, or \emph{geometrically translating}, a
formula to a different location.  For example, translating the
formula \texttt{=SUM(A1:B1)} from cell \texttt{C1} to \texttt{C2}
results in the formula \texttt{=SUM(A2:B2)} (i.e., references are
updated). \excelint{} encodes every reference in a spreadsheet as a reference vector, including references to other worksheets and workbooks. We describe
the form of a reference vector below.

Formally, let $f_1$ and $f_2$ denote two formulas, and let $v$ denote the function that induces a set of reference vectors from a formula.

\begin{lemma}
\label{thm:equivalence}
$f_1$ and $f_2$ are reference-equivalent if and only if $v(f_1) = v(f_2)$.
\end{lemma}

This property is intuitively true: no two formulas can be ``the same'' if they refer to different relative data offsets.  In the base case, $f_1$ and $f_1$ are trivially reference-equivalent.  Inductively, $f_1$ and $f_2$ (where $f_1 \neq f_2$) are reference-equivalent if there exists a translation function $t$ such that $f_2 = t(f_1)$.  Since reference vectors abstract over translation, $v(f_1) = v(f_2)$; therefore, reference equivalence also holds for the transitive closure of a given translation.

\paragraph{Reference vector encoding:}
\begin{sloppypar}
	Reference vectors have the form $\mbox{v} = (\Delta x, \Delta y, \Delta z, \Delta c)$ where $\Delta x$, $\Delta y$, and $\Delta z$ denote numerical column, row, and worksheet offsets with respect to a given \emph{origin}.  The origin for $\Delta x$ and $\Delta y$ coordinates depends on their addressing mode (see below).  $\Delta z$ is 0 if a reference points on-sheet and 1 if it points off-sheet (to another sheet).  $\Delta c$ is 1 if a constant is present, 0 if it is absent, or $-1$ if the cell contains string data.\end{sloppypar}

The entire dataflow graph of a spreadsheet is encoded in vector form.  Since numbers, strings, and whitespace refer to nothing, numeric, string, and whitespace cells are encoded as degenerate \emph{null vectors}.  The $\Delta x$, $\Delta y$, and $\Delta z$ components of the null vector are zero, but $\Delta c$ may take on a value depending on the presence of constants or strings.

\paragraph{Addressing modes:} Spreadsheets have two \emph{addressing modes}, known as \emph{relative addressing} and \emph{absolute addressing}. For example, the reference \texttt{\$A1} has an absolute horizontal and a relative vertical component while the reference \texttt{A\$1} has a relative horizontal and an absolute vertical component. 

  In our encoding, these two modes differ with respect to their origin.  In relative addressing mode, an address is an offset from a formula.  In absolute addressing mode, an address is an offset from the top left corner of the spreadsheet.  The horizontal and vertical components of a reference may mix addressing modes.  
  
Addressing modes are not useful by themselves.  Instead, they are annotations that help Excel's automated copy-and-paste tool, called Formula Fill, to update references for copied formulas. Copying cells using Formula Fill does not change their absolute references. Failing to correctly employ reference mode annotations causes Formula Fill to generate incorrect formulas. Separately encoding these references helps find these errors.

\subsection{Computing the Vector-Based IR}
\label{sec:vectorcompute}

The transformation from formulas to the vector-based internal
representation starts by building a dataflow graph for the
spreadsheet. Each cell in the spreadsheet is represented by a single
vertex in the dataflow graph, and there is an edge for every
functional dependency.  Since spreadsheet expressions are purely
functional, dependence analysis yields a DAG.

\begin{wrapfigure}[14]{r}{0.5\textwidth}
  \begin{center}
    \includegraphics[width=\linewidth]{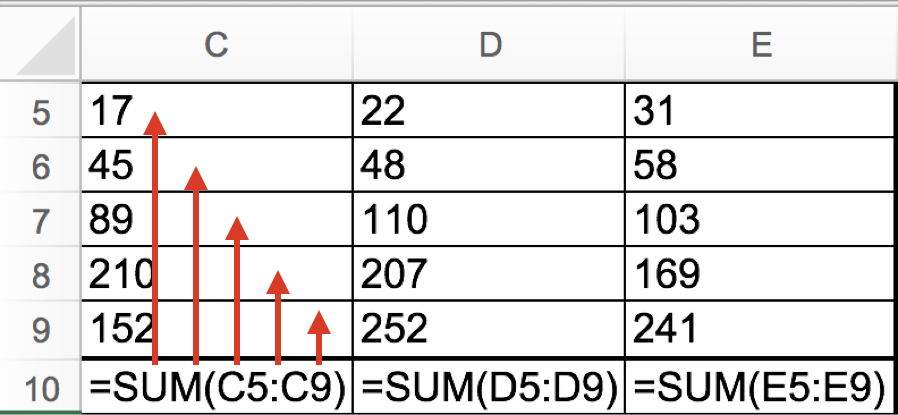}
  \end{center}
  \caption{\textbf{Reference vectors.} The formula in cell \texttt{C10} ``points'' to data in cells \texttt{C5:C9}.  The cell's set of reference vectors are shown in red.  One such vector, representing \texttt{C10}$\rightarrow$\texttt{C5}, is $(0,5,0,0)$.\label{fig:references}}
\end{wrapfigure}

One fingerprint vector is produced for every cell in a spreadsheet,
whether it is a formula, a dependency, or an unused cell. The
algorithm first uses the dependency graph to identify each cell's
immediate dependencies. Next, it converts each dependency to a
reference vector.  Finally, it summarizes the cell with a reference
fingerprint.

For instance, the cell shown in Figure~\ref{fig:references},
\texttt{C10}, uses Excel's ``range'' syntax to concisely specify a
dependence on five inputs, \texttt{C5}$\ldots$\texttt{C9} (inclusive).
As each reference is relatively addressed, the base offset for the
address is the address of the formula itself, \texttt{C10}. There are
no off-sheet references and the formula contains no constants, so the
formula is transformed into the following set of reference vectors:
$\{(0,-5,0,0)\dots(0,-1,0,0)\}$.  After summing, the fingerprint vector
for the formula is $(0,-15,0,0)$.

A key property of fingerprint vectors is that other formulas with the
same reference ``shape'' have the same fingerprint vectors.  For
example, \texttt{=SUM(D5:D9)} in cell \texttt{D10} also has the
fingerprint $(0,-15,0,0)$.

\subsection{Performing Rectangular Decomposition}
\label{sec:decomp}

\begin{wrapfigure}{r}{0.50\textwidth}
  \begin{centering}
    \begin{codebox}
	\Procname{$\proc{EntropyTree}(S)$} \li \If |S| = 1 \Then \li
        \Return $\proc{Leaf}(S)$ \li \Else \li $(l, t, r, b) \gets S$
        \li $x \gets \argmin_{l \leq i \leq r}
        \proc{Entropy}(S,i,\mbox{true})$ \li $y \gets \argmin_{t \leq
          y \leq b} \proc{Entropy}(S,i,\mbox{false})$ \li $p_1 \gets
        (l,t,r,y)$ ; $p_2 \gets (l,y,r,b)$ \li $e \gets
        \proc{Entropy}(S,y,\mbox{false})$ \li \If
        $\proc{Entropy}(S,x,\mbox{true}) \leq
        \proc{Entropy}(S,y,\mbox{false})$ \Then \li $p_1 \gets
        (l,t,x,b)$ ; $p_2 \gets (x,t,r,b)$ \li $e \gets
        \proc{Entropy}(S,x,\mbox{true})$
	      \End
	\li   \If $e = 0.0$ and $\proc{Values}(p_1) = \proc{Values}(p_2)$ \Then
	\li     \Return $\proc{Leaf}(S)$
	\li   \Else
	\li     $t_1 \gets \proc{EntropyTree}(p1)$
	\li     $t_2 \gets \proc{EntropyTree}(p2)$
	\li     \Return $\proc{Node}(t_1, t_2)$
	      \End
	    \End
	\end{codebox}
  \end{centering}
  \caption{$\proc{EntropyTree}$ decomposes a spreadsheet into a tree of rectangular regions by minimizing the sum total entropy vector fingerprint distributions across splits. See \S\ref{sec:decomp} for definitions.}
  \label{fig:entropy-decomp}
\end{wrapfigure}

As noted in the overview, spreadsheet user interfaces strongly encourage users to organize
their data into rectangular shapes. Because built-in spreadsheet operations
expect rectangular layouts, users who avoid them later discover that their data is difficult to manipulate~\cite{Barowy:2015:FER:2737924.2737952}. As a
result, formulas that perform the same operation (and are therefore reference-equivalent) are often
placed in the same row or column.

Since slight deviations in references along a row or column strongly
suggest errors, the rectangular decomposition algorithm aims to divide
a spreadsheet into the user's intended rectangular regions and thus reveal
deviations from them. To produce regions that faithfully capture user intent,
our algorithm generates a rectangular decomposition with the following
properties.

First, since users often use strings and whitespace in semantically meaningful ways, such as dividers between different computations, every cell, whether it contains a string, whitespace, a number, or a formula, must belong to exactly one rectangular region.

Second, rectangular regions should be as large as possible while maintaining the property that all of the cells in that region have the same fingerprint vector.


Our rectangular decomposition algorithm performs a top-down, recursive
decomposition that splits spreadsheet regions into subdivisions by
minimizing the \emph{normalized
  Shannon entropy}~\cite{shannon}, an information-theoretic statistic.  Specifically, it performs a
recursive binary decomposition that, at each step, chooses the split
that minimizes the sum of the normalized Shannon entropy of vector
fingerprints in both subdivisions. We use normalized Shannon entropy
since the binary partitioning process does not guarantee that entropy
comparisons are made between equal-sized sets; normalization
ensures that comparisons are well-behaved~\cite{GEAN:GEAN1014}.

Normalized Shannon entropy is defined as:

\[
	\eta(X) = -\sum_{i=1}^n \frac{p(x_i) \log_b{p(x_i)}}{\log_b{n}}
\]

where $X$ is a random vector denoting cell counts for each fingerprint region,  where $x_i$ is a given fingerprint count, where $p(x_i)$ is the count represented as a probability, and where $n$ is the total number of cells.  Large values of $\eta$ correspond with complex layouts, whereas small values correspond to simple ones.  When there is only one region, $\eta$ is defined as zero.

The procedure $\proc{EntropyTree}$ in Figure~\ref{fig:entropy-decomp}
presents the recursive rectangular decomposition algorithm.  The
algorithm returns a binary tree of regions, where a region is a
4-tuple consisting of the coordinates \texttt{(left, top, right,
  bottom)}.  $S$ is initially the entire spreadsheet.  Each region
contains only those cells with exactly the same fingerprint.

$\proc{Entropy}$ computes the normalized Shannon entropy of
spreadsheet $S$ along the split $i$ which is an $x$ coordinate if $v =
\mbox{true}$, otherwise the coordinate is $y$ (i.e., $v$ controls whether a split is horizontal or vertical).  $p1$ and $p2$ represent the rectangles induced by a partition. The normalized entropy of the
empty set is defined as $+\infty$. $\proc{Values}$ returns the set of
distinct fingerprint vectors for a given region.  Finally,
$\proc{Leaf}$ and $\proc{Node}$ are constructors for a leaf tree node
and an inner tree node, respectively.


$\proc{EntropyTree}$ is inspired by the \idthree decision tree
induction algorithm~\cite{Quinlan86inductionof}.
As with \idthree, $\proc{EntropyTree}$ usually produces a good binary
tree, although not necessarily the optimally compact one.  Instead,
the tree is decomposed greedily.  In the worst case,
the algorithm places each cell in its own subdivision.  To have arrived at this
worst case decomposition, the algorithm would have computed the entropy for
all other rectangular decompositions first.  For a grid of height $h$ and width $w$,
there are $\frac{h^2w^2 + h^2w + hw^2 + hw}{4}$ possible rectangles, so entropy is computed $O(h^2w^2)$ times.

\begin{sloppypar}Finally, regions for a given spreadsheet are obtained by running $\proc{EntropyTree}$  and extracting them from the leaves of the returned tree.	
\end{sloppypar}

\paragraph{Adjacency coalescing:} $\proc{EntropyTree}$ sometimes produces two or more adjacent regions containing the same fingerprint. Greedy decomposition does not usually produce a globally optimal decomposition; rather it chooses local minima at each step.  \emph{Coalescing} merges pairs of regions subject to two rules, producing better regions: (1) regions are adjacent, and (2) the merge is a contiguous, rectangular region of cells.

Coalescing is a fixed-point computation, merging two regions at every
step, terminating when no more merges are possible.  In the worst
case, this algorithm takes time proportional to the total number of
regions returned by $\proc{EntropyTree}$.  In practice, the algorithm
 terminates quickly because the binary tree is close to the
ideal decomposition.

\subsection{Proposed Fix Algorithm}

\label{sec:ProposedFix}

When a user fixes a formula error, that formula's fingerprint vector changes. The purpose of the proposed fix algorithm is to explore the effects of
such fixes.  A \emph{proposed fix} is an operation that mimics the
effect of ``correcting'' a formula.  The working hypothesis of the
analysis is that surprising regions are likely wrong, and that
unsurprising regions are likely correct.  \excelint{}'s analysis
leverages this fact to identify which cells should be fixed, and how.
Those fixes that do not cause formulas to merge with existing regions
are not likely to be good fixes.


Formally, a proposed fix is the tuple $(s,t)$, where $s$ is a
(nonempty) set of \emph{source cells} and $t$ is a (nonempty) set of
\emph{target cells}.  $t$ must always be an existing region but $s$
may not be; source cells may be borrowed from other regions.  A
proposed fix should be thought of as an operation that \emph{replaces
  the fingerprints} of cells in $s$ with the fingerprint of cells in
$t$.
%

\subsubsection{Entropy-Based Error Model}
\label{sec:EntropyModel}

Not all proposed fixes are good, and some are likely bad. \excelint{}'s
static analysis uses an error model to identify which fixes are the
most promising.  A good model helps users to identify errors and to assess the impact of correcting them.

We employ an entropy-based model. Intuitively, formula errors result
in irregularities in the set of rectangular regions and so increase
entropy \emph{relative to the same spreadsheet without errors}. A
proposed fix that reduces entropy may thus be a good fix because it
moves the erroneous spreadsheet closer to the correct spreadsheet.

Since most formulas belong to large rectangular regions (\sectref{sec:rq0}), many formulas outside those regions are likely errors (\sectref{sec:rq1}).  The entropy model lets the analysis explore the impact of fixing these errors---making the spreadsheet
more regular---by choosing only the most promising ones which are then
presented to the user.

Formally, $m$ is a set of rectangular regions.  A set of proposed
fixes of size $n$ yields a set of new spreadsheets $m'_1 \ldots m'_n$, where
each $m'_i$ is the result of one proposed fix $(s,t)_i$.  The
\emph{impact} of fix the $(s,t)_i$ is defined as the difference in
normalized Shannon entropy, $\delta \eta_i = \eta(m_i) - \eta(m)$.

Positive values of $\delta \eta_i$ correspond to increases in entropy
and suggest that a proposed fix is bad because the spreadsheet has
become more irregular.  Negative values of $\delta \eta_i$ correspond
to decreases in entropy and suggest that a proposed fix is good.

Somewhat counterintuitively, fixes that result in large decreases in
entropy are worse than fixes that result in small decreases.  A fix
that changes large swaths of a spreadsheet will result in a large
decrease in entropy, but this large-scale change is not necessarily a good fix for
several reasons.  First, we expect bugs to make up only a small
proportion of a spreadsheet, so fixing them should result in
\emph{small} (but non-zero) decreases in entropy.  The best fixes are
those where the prevailing reference shape is a strong signal, so
corrections are minor.  Second, large fixes are more work.  An
important goal of any bug finder is to minimize user effort. Our
approach therefore steers users toward those hard-to-find likely
errors that minimize the effort needed to fix them.


\subsubsection{Producing a Set of Fixes}

The proposed fix generator then considers all fixes for every possible
source $s$ and target $t$ region pair in the spreadsheet.  A na\"ive
pairing would likely propose more fixes than the user would want to
see.  In some cases, there are also more fixes than the likely number
of bugs in the spreadsheet.  Some fixes are not independent; for
instance, it is possible to propose more than one fix utilizing the
same source region.  Clearly, it is not possible to perform both
fixes.

As a result, the analysis suppresses certain fixes, subject to the
conditions described below.  These conditions are not heuristic in nature; rather, they address conditions that naturally arise when considering the kind of dependence structures that can arise when laid out in a 2D grid. The remaining fixes are scored by a
fitness function, ranked from most to least promising, and then
thresholded.  All of the top-ranked fixes above the threshold are
returned to the user.

The cutoff threshold is a user-defined parameter that represents the
proportion of the worksheet that a user is willing to inspect.  The
default value, 5\%, is based on the observed frequency of spreadsheet
errors in the
wild~\cite{PankoEUSPRIG2015,Barowy:2014:CDD:2660193.2660207}.  Users
may adjust this threshold to inspect more or fewer cells, depending on
their preference.

\paragraph{Condition 1: Rectangularity:}

Fixes must produce rectangular layouts.  This condition arises from
the fact that Excel and other spreadsheet languages have many
affordances for rectangular composition of functions.

\paragraph{Condition 2: Compatible Datatypes:}

Likely errors are those identified by fixes $m_i$ that produce small,
negative values of $\delta \eta_i$.  Nonetheless, this is not a
sufficient condition to identify an error.  Small regions can belong
to data of any type (string data, numeric data, whitespace, and other
formulas).  Fixes between regions of certain datatypes are not likely
to produce desirable effects.  For instance, while a string may be
replaced with whitespace and vice-versa, neither of these proposed
fixes have any effect on the computation itself. We therefore only consider fixes where both the source and target regions are formulas.


\paragraph{Condition 3: Inherently Irregular Computations:}

A common computational structure is inherently unusual in the
spreadsheet domain: aggregates.

%

An \emph{aggregate} consists of a formula cell $f_a$ and a set of
input cells $c_0, \ldots, c_n$ such that $f_a$ refers exclusively to
$c_0, \ldots, c_n$.  This computational form is often seen in
functional languages, particularly in languages with a statistical
flavor.  Excel comes with a large set of built-in statistical
functions, so functions of this sort are invoked frequently.  Examples
are the built-in functions \texttt{SUM} and \texttt{AVERAGE}.  $f_a$
often sits adjacent to $c_0, \ldots, c_n$, and so $f_a$ appears
surprising when compared to $c_0, \ldots, c_n$.  Without correction, $f_a$ would frequently rank highly as a likely error.  Since aggregates are usually false positives, no fix is proposed for formulas of this form.

Note that this restriction does not usually impair the ability
of the analysis to find errors.  The reason is that the relevant comparison is not between an aggregate formula and its inputs, but between an aggregate formula and \emph{adjacent aggregate formulas}.

For example, the analysis can still find an off-by-one error, where an aggregate like \texttt{SUM}
refers to either \emph{one more} or \emph{one fewer} element.  Let $b_0, \ldots, b_n$ be a rectangular region, and let $f_b$ be a formula that incorrectly aggregates those cells.  Let $c_0, \ldots, c_n$ be a rectangular region adjacent to  $b_0, \ldots, b_n$, and let $f_c$ be a formula adjacent to $f_b$ that \emph{correctly} aggregates $c_0, \ldots, c_n$.  Then  if $f_b$ refers to $b_0, \ldots, b_n, d$ (one more) or $b_0, \ldots, b_{n-1}$ (one fewer), $f_b$ is a likely error.  Because the analysis only excludes proposed fixes for aggregates that refer to $c_0, \ldots, c_n$, we still find the error since $f_b$ will eventually be compared against $f_c$.

\subsubsection{Ranking Proposed Fixes}

After a set of candidate fixes is generated, \excelint's analysis ranks them according to an impact score.  We first formalize a notion of ``fix distance,'' which is a measure of the similarity of the references of two rectangular regions.  We then define an impact score, which allows us to find the ``closest fix'' that also causes small drops in entropy.

\paragraph{Fix distance:}

Among fixes with an equivalent entropy reduction, some fixes are
better than others.  For instance, when copying and pasting formulas,
failing to update one reference is more likely than failing to update
all of them, since the latter has a more noticeable effect on the
computation.  Therefore, a desirable criterion is to favor smaller
fixes using a location-sensitive variant of vector fingerprints.

We use the following distance metric, inspired by the earth
mover's distance~\cite{M81}:

\[
  d(x,y) = \sum_{i=1}^n \sqrt{ \sum_{j=1}^k ( h_s(x_i)_j - h_s(y_i)_j )^2}
\]

where $x$ and $y$ are two spreadsheets, where $n$ is the number of cells in
both $x$ and $y$, where $h_s$ is a location-sensitive fingerprint hash
function, where $i$ indexes over the same cells in both $x$ and $y$,
and where $j$ indexes over the vector components of a fingerprint
vector for fingerprints of length $k$.  The intuition is that formulas with small errors are more likely to escape the notice of the programmer than large swaths of formulas with errors, thus errors of this kind are more likely left behind. Since we model formulas as clusters of references, each reference represented as a point in space, then we can measure the ``work of a fix'' by measuring the cumulative distance it takes to ``move'' a given formula's points to make an erroneous formula look like a ``fixed'' one.  Fixes that require a lot of work are ranked lower.

\paragraph{Entropy reduction impact score:}

The desirability of a fix is determined by an \emph{entropy reduction
  impact score}, $S_i$. $S_i$ computes the potential improvement between the original spreadsheet, $m$, and the fixed spreadsheet, $m_i$. As a shorthand, we use $d_i$ to refer to the distance $d(m,m_i)$.

\[
  S_i = \frac{n_t}{-\delta\eta_i d_i}
\]

where $n_t$ is the size of the target region, $\delta\eta_i$ is
difference in entropy from $m_i$ to $m$, and $d$ is the fix distance.

Since the best fixes minimize $-\delta\eta_i$, such fixes maximize $S_i$.  Likewise, ``closer'' fixes according to the distance metric also produce higher values of $S_i$.  Finally, the score leads to a preference for fixes whose ``target'' is a large region.  This preference ensures that the highest ranked deviations are actually rare with respect to a reference shape.




\section{\excelint{} Implementation}
\label{sec:implementation}
\excelint is written in C\# and F\# for the .NET managed
language runtime, and runs as a plugin for Microsoft Excel (versions
2010-2016) using the Visual Studio Tools for Office framework.
We first describe key optimizations in \excelint's implementation (\S\ref{sec:optimizations}), and then discuss \excelint's visualizations (\S\ref{sec:visualizations}).

\subsection{Optimizations}
\label{sec:optimizations}

Building an analysis framework to provide an interactive level of
performance was a challenging technical problem during \excelint's
development.  Users tend to have a low tolerance for tools that make
them wait. This section describes performance
optimizations undertaken to make \excelint fast. Together, these
optimizations produced orders of magnitude improvements
in \excelint's running time.

\subsubsection{Reference Fingerprints}
\label{sec:fingerprintoptim}

Reference vector set comparisons are the basis for the inferences made
by \excelint's static analysis algorithm.  The cost of comparing a vector set is the cost of comparing two vectors times the cost of set comparison.  While set comparison can be made reasonably fast (e.g., using the union-find data structure), \excelint utilizes an even faster approximate data structure that allows for constant-time comparisons.  We call this approximate data structure a \emph{reference fingerprint}.  Fingerprint comparisons are computationally inexpensive and can be performed liberally throughout the analysis.



\paragraph{Definition:} A vector fingerprint summarizes a formula's set of reference vectors.  Let $f$ denote a formula and $v$ a function that induces reference vectors from a formula.  A vector fingerprint is the hash function: $h(f) = \sum_{i \in v(f)} i$ where $\sum$ denotes vector sum.

%

When two functions $x$ and $y$ with disjoint reference vector sets
$r(x)$ and $r(y)$ have the same fingerprint, we say that they
\emph{alias}. For example, the fingerprint $(-3,0,0,0)$ is induced
both by the formula \texttt{=SUM(A1:B1)} in cell \texttt{C1} and the
formula \texttt{=ABS(A1)} in cell \texttt{D1}, so the two formulas
alias.  Therefore, Lemma~\ref{thm:equivalence} does not hold for fingerprints.  Specifically, only one direction of the relation holds: while it is true that two formulas with different fingerprints are guaranteed not to be reference-equivalent, the converse is not true.

Fortunately, the likelihood of aliasing $P[h(f_1)
  = h(f_2) \land v(f_1) \neq v(f_2)]$ is small for fingerprints, and thus the property holds with high probability.  Across the spreadsheet corpus used in our benchmarks (see Section~\ref{sec:evaluation}), on average, \meanHashCollisionsPct of fingerprints in a workbook collide (median collisions per workbook = \medianHashCollisionsPct).

The low frequency of aliasing justifies the use of fingerprints compared to exact formula comparisons.  \excelint's analysis also correctly concludes that expressions like \texttt{=A1+A2} and \texttt{=A2+A1} have the same reference behavior; comparisons like this still hold with the approximate version.

\subsubsection{Grid Preprocessing Optimization}
\label{sec:gridpreprocessing}

One downside to the $\proc{EntropyTree}$ algorithm described in
Section~\ref{sec:algorithms} is that it can take a long time on large
spreadsheets.  While spreadsheets rarely approach the maximum size
supported in Microsoft Excel (16,000 columns by 1,000,000 rows),
spreadsheets with hundreds of rows and thousands of columns are not
unusual. $\proc{EntropyTree}$ is difficult to parallelize because
binary splits rarely contain equal-sized subdivisions, meaning that
parallel workloads are imbalanced.

Nonetheless, one can take advantage of an idiosyncrasy in the way that
people typically construct spreadsheets to dramatically speed up this
computation.  People frequently use contiguous, through-spreadsheet
columns or rows of a single kind of value as delimiters.  For example,
users often separate a set of cells from another set of cells using
whitespace.

\begin{sloppypar}
	By scanning the spreadsheet for through-spreadsheet columns or rows of equal fingerprints, the optimization supplies the rectangular decomposition algorithm with smaller sub-spreadsheets which it decomposes in parallel.  Regions never cross through-spreadsheet delimiters, so preprocessing a spreadsheet does not change the outcome of the analysis.
\end{sloppypar}

In our experiments, the effect of this optimization was dramatic:
after preprocessing, performing static analysis on large
spreadsheets went \emph{from taking tens of minutes to seconds}.  Scanning
for splits is also inexpensive, since there are only
$O($\mbox{width}$+$\mbox{height}$)$ possible splits.  \excelint uses
all of the splits that it finds.

\subsubsection{Compressed Vector Representation}

In practice, subdividing a set of cells and computing their entropy is
somewhat expensive.  A cell address object in \excelint stores not
just information relating to its $x$ and $y$ coordinates, but also its
worksheet, workbook, and full path on disk.  Each object contains two
32-bit integers, and three 64-bit managed references and is therefore
``big''.  A typical analysis compares tens or hundreds of thousands of
addresses, one for each cell in an analysis.  Furthermore, a
fingerprint value for a given address must be repeatedly recalled or
computed and then counted to compute entropy.

Another way of storing information about the distribution of
fingerprints on a worksheet uses the following encoding, inspired by the optimization used in \flashrelate~\cite{Barowy:2015:FER:2737924.2737952}.  In this scheme, no more than $f$ bits are stored for each address, where $f$ is the number of unique fingerprints.  $f$ is often small, so the total number of bitvectors stored is also small.  The insight is that the number of fingerprints is small relative to the number of cells on a spreadsheet.

For each
unique fingerprint on a sheet, \excelint stores one bitvector.  Every
cell on a sheet is given exactly one bit, and its position in the
bitvector is determined by a traversal of the spreadsheet.  A bit at a
given bit position in the bitvector signifies whether the
corresponding cell has that fingerprint: $1$ if it does, $0$ if not.

The following bijective function maps $(x,y)$ coordinates to a bitvector index: $\mbox{Index}_s(x,y) = (y - 1) \cdot w_s + x - 1$ where $w_s$ is the width of worksheet $s$.  The relation subtracts one from the result because bitvector indices range over $0 \ldots n-1$ while address coordinates range over $1 \ldots n$.

Since the rectangular decomposition algorithm needs to compute the entropy for subdivisions of a worksheet, the optimization needs a low-cost method of excluding cells.  \excelint computes masked bitvectors to accomplish this. The bitvector mask corresponds to the region of interest, where $1$ represents a value inside the region and $0$ represents a value outside the region.  A bitwise \texttt{AND} of the fingerprint bitvector and the mask yields the appropriate bitvector.  The entropy of subdivisions can then be computed, since all instances of a fingerprint appearing outside the region of interest appear as $0$ in the subdivided bitvector.

With this optimization, computing entropy for a spreadsheet largely
reduces to counting the number of ones present in each bitvector,
which can be done in $O(b)$ time, where $b$ is the number of bits
set~\cite{Wegner:1960:TCO:367236.367286}.  Since the time cost of
setting bits for each bitvector is $O(b)$ and bitwise \texttt{AND} is
$O(1)$, the total time complexity is $O(f\cdot b)$, where $f$ is the
number of fingerprints on a worksheet. Counting this way speeds up the analysis by approximately $4\times$.

%

\subsection{Visualizations}
\label{sec:visualizations}

\begin{figure*}[!t]
\centering
\begin{subfigure}{0.50\linewidth}
  \centering
  \includegraphics[width=0.97\linewidth]{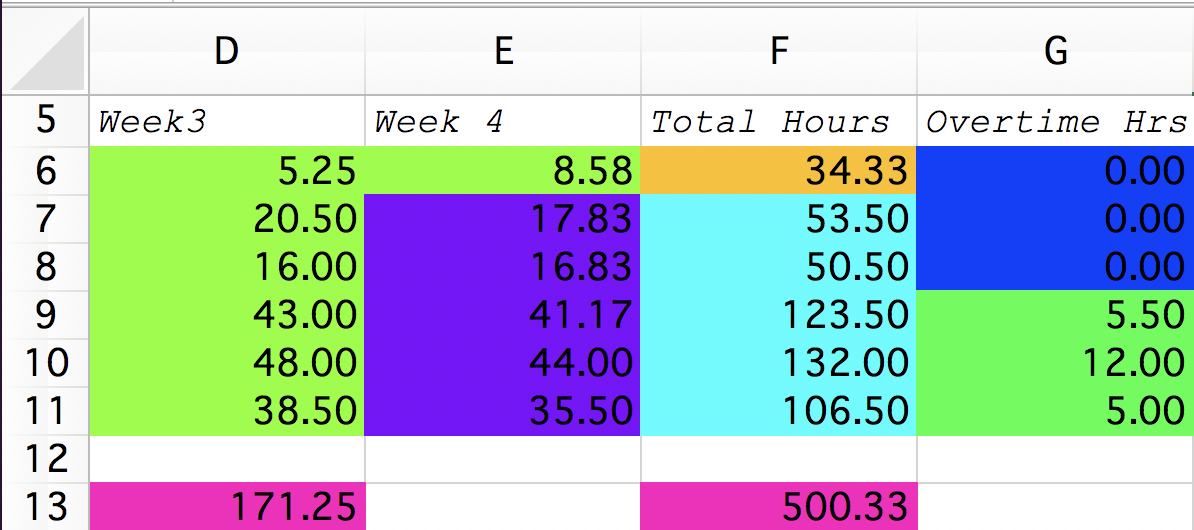}
  \caption{}
  \label{fig:compl}
\end{subfigure}%
\begin{subfigure}{0.50\linewidth}
  \centering
  \includegraphics[width=0.97\linewidth]{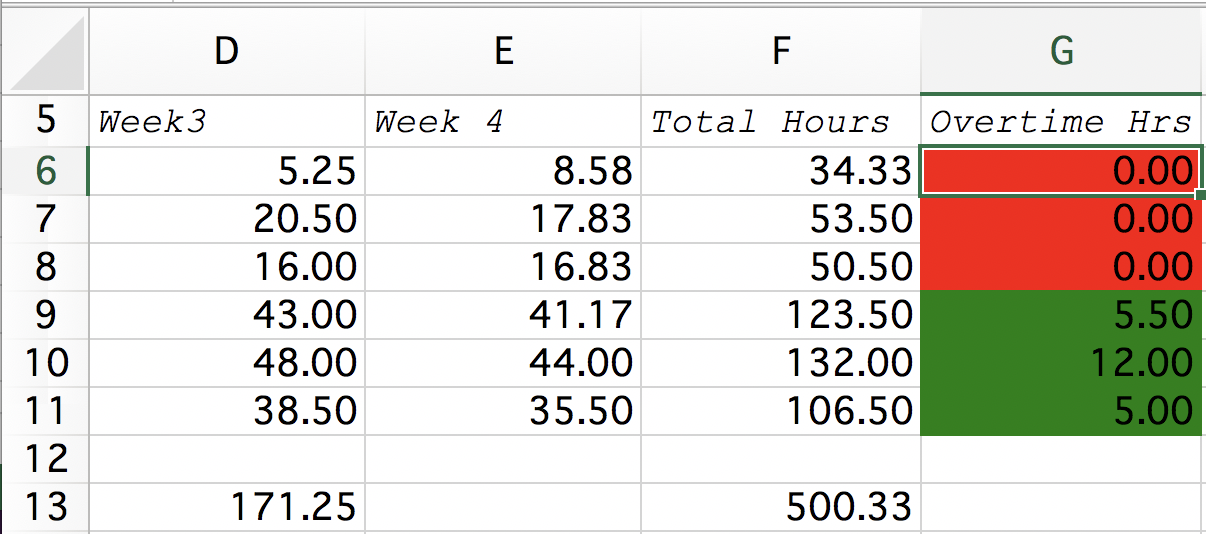}
  \caption{}
  \label{fig:act3flag2}
\end{subfigure}
\caption{\textbf{Global view}: (a) In the global view, colors are mapped to fingerprints so that users equate color with reference equivalence. For example, the block of light green cells on the left are data; other colored blocks represent distinct sets of formulas.  Cells \texttt{G6:G8}, for example, are erroneous because they incorrectly compute overtime using only hours from Week 1. \textbf{Guided audit}: (b) \excelint's guided audit tool flags  \texttt{G6:G8} (red) and suggests which reference behavior should have been used (\texttt{G9:G11}, in green).
}
\label{fig:colorchoice}
\end{figure*}


\excelint provides two visualizations that assist users to find bugs:
the \emph{global view} (\S\ref{sec:GlobalView}) and the \emph{guided audit} (\S\ref{sec:AuditTool}).  Both tools are based on \excelint{}'s
underlying static analysis.


\subsubsection{Global View}
\label{sec:GlobalView}



The \emph{global view} is a visualization for finding potential errors
in spreadsheets.  The view takes advantage of the keen human ability
to quickly spot deviations in visual patterns.  Another example of \excelint's global view the the running example from Figure~\ref{fig:Act3} is
shown in Figure~\ref{fig:colorchoice}. The goal of the global view is to draw attention to irregularities in
the spreadsheet.  Each colored block represents a contiguous region
containing the same formula reference behavior (i.e., where all
cells have the same fingerprint vector).




\afterpage{\FloatBarrier}

While the underlying decomposition is strictly rectangular for the
purposes of entropy modeling, the global view uses the same color in
its visualization anywhere the same vector fingerprint is found.  For
example, all the numeric data in the visualization are shown using the
same shade of blue, even though each cell may technically belong to a
different rectangular region (see Figure~\ref{fig:compl}). This
scheme encourages users to equate color with reference
behavior. Whitespace and string regions are not colored in the
visualization to avoid distracting the user.

\begin{sloppypar}The global view chooses colors specifically to maximize perceptual differences (see Figure~\ref{fig:compl}).  Colors are assigned such that adjacent clusters use complementary or near-complementary colors.  To maximize color differences, we use as few colors as possible. This problem corresponds exactly to the classic graph coloring problem.	
\end{sloppypar}

The color assignment algorithm works by building a graph of all
adjacent regions, then colors them using a greedy coloring heuristic
called largest degree ordering~\cite{doi:10.1093/comjnl/10.1.85}.
This scheme does not produce the optimally minimal coloring, but it does have the
benefit of running in $O(n)$ time, where $n$ is the number of vertices
in the graph.


Colors are represented internally
using the Hue-Saturation-Luminosity (HSL) model, which models the space of colors as a cylinder.
The cross-section of a cylinder is a circle; hue corresponds to the
angle around this circle.  Saturation is a number from 0 to 1 and
represents a point along the circle's radius, zero being at the
center.  Luminosity is a number between 0 and 1 and represents a point
along the length of the cylinder.

New colors are chosen as follows.  The algorithm starts by choosing colors at the starting point of hue =
$180^\circ$, saturation 1.0, and luminosity 0.5, which is bright blue.
Subsequent colors are chosen with the saturation and luminosity fixed,
but with the hue being the value that maximizes the distance on the
hue circle between the previous color and any other color.  For
example, the next color would be HSL($0^\circ$, 1.0, 0.5) followed by HSL($90^\circ$, 1.0, 0.5).  The algorithm is also parameterized by
a color restriction so that colors may be excluded for other purposes.
For instance, our algorithm currently omits bright red, a color commonly associated with other errors or warnings.


\afterpage{\FloatBarrier}

\subsubsection{Guided Audit}
\label{sec:AuditTool}


Another visualization, the \emph{guided audit}, automates some of
the human intuition that makes the global view effective.
This visualization is a cell-by-cell audit of the highest-ranked
proposed fixes generated in the third phase of the
excelint{}' static
analysis described in Section~\ref{sec:EntropyModel}.
Figure~\ref{fig:act3flag2} shows a sample fix.  The portion in
red represents a set of potential errors, and the portion in green represents
the set of formulas that \excelint thinks maintains the correct
behavior.  This fix a good suggestion, as the formulas
in \texttt{G6:G8} incorrectly omit data when computing overtime.



While we often found ourselves consulting the global view for
additional context with respect to cells flagged by the guided audit,
the latter is critical in an important scenario:
large spreadsheets that do not fit on-screen.  The guided audit solves this scalability issue by highlighting only one suspicious region at a time.  The analysis also highlights the region likely to correctly observe the intended reference behavior.

When a user clicks the ``Audit'' button, the guided audit
highlights and centers the user's window on each error, one at a time.
To obtain the next error report, the user clicks the ``Next Error''
button.  Errors are visited according to a ranking from most to least
likely.  Users can stop or restart the analysis from the beginning at
any time by clicking the ``Start Over'' button.  If \excelint{}'s
static analysis reports no bugs, the guided audit highlights
nothing. Instead, it reports to the user that the analysis found no
errors.  For performance reasons, \excelint{} runs the analysis only once; thus the analysis is not affected by corrections the user makes during an audit.

\section{Evaluation}
\label{sec:evaluation}

\begin{wrapfigure}[13]{r}{0.45\textwidth}
\vspace{-1em}
  \begin{centering}
    \includegraphics[width=\linewidth]{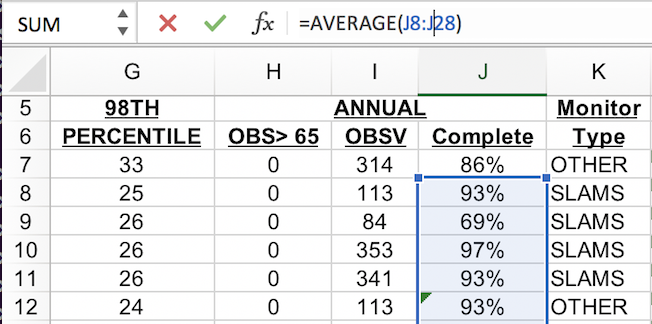}
  \end{centering}
  \caption{\textbf{Example error found by \excelint{}:} the formula shown (in cell \texttt{J30}) has an off-by-one error that omits a row (from {\tt 01sumdat.xls}, sheet {\tt PM2.5}).}
  \label{fig:ExampleRefError}
\end{wrapfigure}

The evaluation of \excelint focuses on answering the following research questions. (1) Are spreadsheet layouts really rectangular?  (2) How does the proposed fix tool compare against a state-of-the-art pattern-based tool used as an error finder? (3) Is \excelint fast enough to use in practice? (4) Does it find known errors in a professionally audited spreadsheet? 

\subsection{Definitions}
\label{sec:evaldefs}

\subsubsection*{Formula Error} We strictly define a \emph{formula error} as a formula that deviates from the intended reference shape by either \emph{including} an extra reference, \emph{omitting} a reference, or \emph{misreferencing} data.  We also include manifestly wrong calculations in this category, such as choosing the wrong operation.  A formula error that omits a reference is shown in Figure~\ref{fig:ExampleRefError}.

\subsubsection*{Inconsistent Formula Pairs}
In our spreadsheet corpus, which we borrowed from the \custodes project, incorrect formulas are often found adjacent to correct formulas.  This paired structure complicates counting errors because for a given pair, it is often impossible to discern which set is correct and which is incorrect without knowing the user's intent.  Despite this, it is usually clear that both sets cannot be correct.

\subsubsection*{Bug Duals}
We call these pairs of inconsistent formula sets \emph{bug duals}.  Formally, a bug dual is a pair containing two sets of cells, $(c_1,c_2)$.  In general, we do not know which set of cells, $c_1$ or $c_2$, is correct.  We do know, however, that all cells in $c_1$ induce one fingerprint and that all cells in $c_2$ induce another.

Without knowing which set in a dual is correct, we cannot count the ``true number of errors.''  We thus arbitrarily label the smaller set of formulas ``the error'' and the larger set ``correct.''  This is more likely to be a good labeling than the converse because errors are usually rare.  Nonetheless, it is occasionally the case that the converse is a better labeling: the user made \emph{systematic errors} and the entire larger region is wrong.  For example, an incautious user can introduce large numbers of errors when copying formulas if they fail to update references appropriately.


Our labeling scheme more closely matches the amount of real work that a user must perform when discovering and fixing an inconsistency.  In the case where \excelint mislabels the pair---i.e., the larger region is in error---investigating the discrepancy still reveals the problem.  Furthermore, the marginal effort required to fix the larger set of errors versus a smaller set of errors is small.  Most of the effort in producing bug fixes is in identifying and understanding an error, not in fixing it, which is often mechanical (e.g., when using tools like Excel's ``formula fill'').  Counting bug duals by the size of the smaller set is thus a more accurate reflection of the actual effort needed to do an audit.

\subsubsection*{Counting Errors}
We therefore count errors as follows.  Let a cell flagged by an error-reporting tool be called a \emph{flagged cell}.  If a flagged cell is not a formula error, we add nothing to the total error count.  If a flagged cell is an error, but has no dual, we add one to the total error count.  If a flagged cell is an error and has a bug dual, then we maintain a count of all the cells flagged for the given dual.  The maximum number of errors added to the total error count is the number of cells flagged from either set in the dual, or the size of the smaller region, whichever number is \emph{smaller}.

\subsection{Evaluation Platform}

All evaluation was performed on a developer workstation, a 3.3GHz Intel Core i9-7900X with a 480GB solid state disk, and 64GB of RAM.  We used the 64-bit version of Microsoft Windows 10 Pro.  We also discuss performance on a more representative platform in \sectref{sec:rq3}.
\afterpage{\FloatBarrier}

\subsection{Ground Truth}
\label{sec:rq1}

In order to evaluate whether \excelint's global view visualization finds new errors, we manually examined \numexsheets publicly-available spreadsheets, using the \excelint global view to provide context.  This manual audit produced a set of annotations which we use for the remainder of the evaluation.

\paragraph{About the benchmarks:} We borrow \numexsheets spreadsheet benchmarks developed by researchers (not us) for a comparable spreadsheet tool called \custodes~\cite{custodes}.  These benchmarks are drawn from the widely-used EUSES corpus~\cite{Fisher:2005:ESC:1082983.1083242}.  These spreadsheets range in size from \SuiteMinCells cells to \SuiteMaxCells cells (mean: \SuiteMeanCells).  The number of formulas for each benchmark ranges from \SuiteMinFormulas to \SuiteMaxFormulas (mean: \SuiteMeanFormulas).  Most spreadsheets are moderate-to-large in size.

The EUSES corpus collects spreadsheets used as databases, and for financial, grading, homework, inventory, and modeling purposes.  EUSES is frequently used by researchers building spreadsheet tools~\cite{Barowy:2015:FER:2737924.2737952,Barowy:2014:CDD:2660193.2660207,custodes,Joharizadeh:2015:FBS:2814189.2815373,Hofer:2013:EEF:2450312.2450321,Hermans:2013:DCD:2486788.2486827,Alawini:2015:TAP:2791347.2791358,melford-using-neural-networks-find-spreadsheet-errors,Hermans:2010:AEC:1883978.1883984,Hermans:2012:DVI:2337223.2337275,Hermans:2014:BRE:2635868.2661673,Grigoreanu:2010:SAD:1753326.1753431,Le:2014:FFD:2594291.2594333,Muslu:2015:PDE:2771783.2771792}.  All of the categories present in EUSES are represented in the \custodes suite.

\sloppypar{Note that during the development of \excelint, we primarily utilized a small number of synthetic benchmarks generated by us, some spreadsheets from the FUSE corpus~\cite{DBLP:conf/msr/BarikLSSM15}, and one spreadsheet (\texttt{act3\_lab23\_posey.xls}) from the \custodes corpus (because it was small and full of errors).}

\paragraph{Procedure:} We re-annotated all of the \numexsheets spreadsheets provided by the \custodes research group.  Re-auditing the same corpus with a different tool helps establish whether \excelint helps uncover errors that \custodes does not find; it also draws a distinction between smells and real formula errors.  The annotation procedure is as follows.  Each spreadsheet was opened and the \excelint global view was displayed.  Visual irregularities were inspected either by clicking on the formula and examining its references, or by using Excel's formula view.  If the cell was found to be a formula error, it was labeled as an error.  In cases where an error had a clear bug dual (see ``Bug duals'' in \sectref{sec:evaldefs}), both sets of cells were labeled, and a note was added that the two were duals.  We then inspected the \custodes ground truth annotations for the same spreadsheet, following the same procedure as before, with one exception: if it was clear that a labeled smell was not a formula error, it was labeled ``not a bug.''

%
%
\paragraph{Results:} For the \numexsheets spreadsheets we annotated, the \custodes ground truth file indicates that \NumSmells cells are smells.  Our audit shows that, among the flagged cells, \custodes finds \CUSTODESNumTrueRefBugs formula errors that also happen to be smells.  During our re-annotation, we found an additional \ECTrueRefBugDelta formula errors, for a total of \ExceLintNumTrueRefBugs formula errors.  We spent a substantial amount of time manually auditing these spreadsheets (roughly \AnnoHours hours, cumulatively).  Since we did not perform an unassisted audit for comparison, we do not know how much time we saved versus not using the tool.  Nonetheless, since an unassisted audit would require examining \emph{all} of the cells individually, the savings are likely substantial.  On average, we uncovered one formula error per \ExceLintMinutesPerBug minutes, which is clearly an effective use of auditor effort.


\begin{figure}[!t]
\centering
   \begin{subfigure}[b]{0.48\linewidth}
   \includegraphics[width=\linewidth]{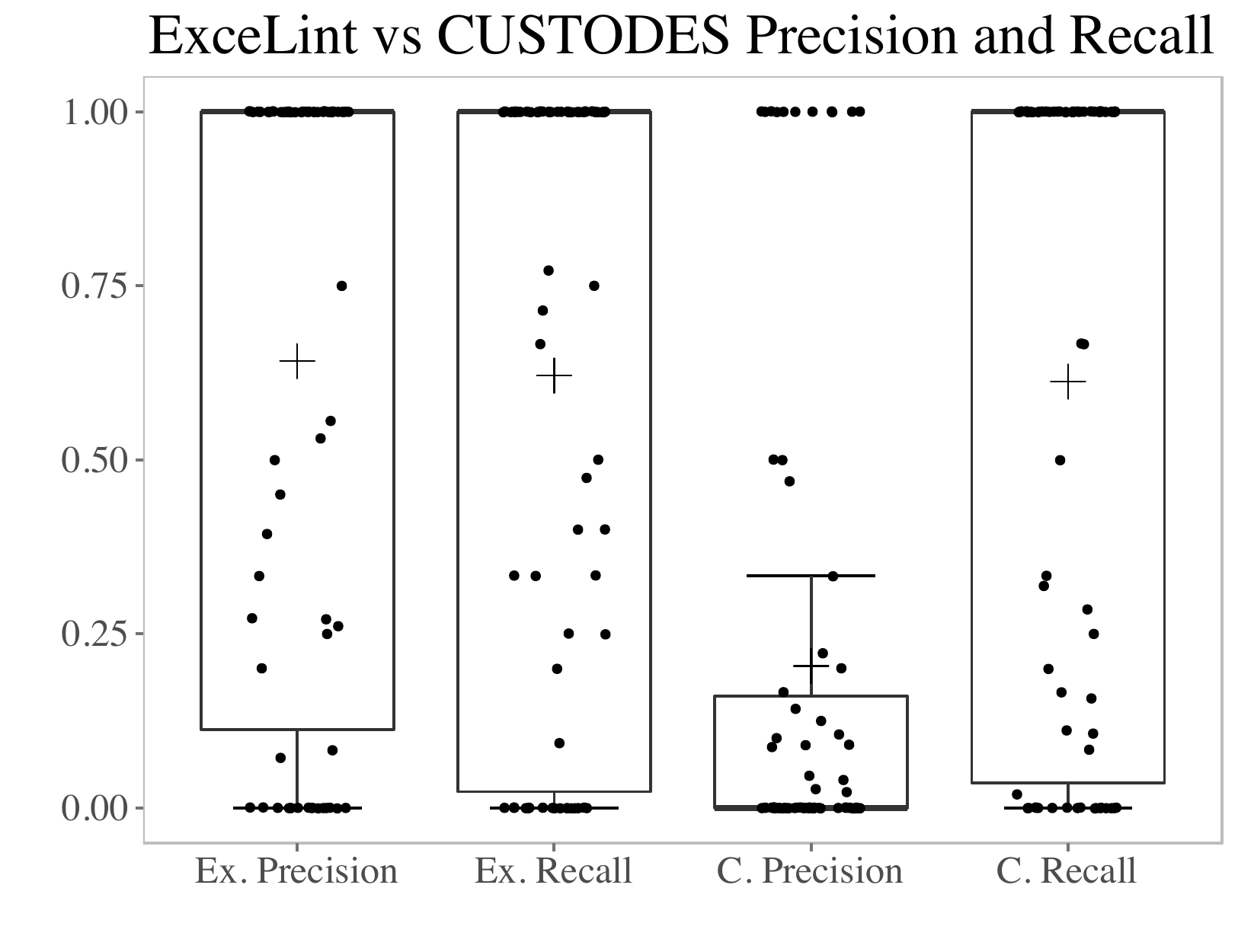}
   \caption{}
   \label{fig:wewin} 
\end{subfigure}
\begin{subfigure}[b]{0.48\linewidth}
   \includegraphics[width=\linewidth]{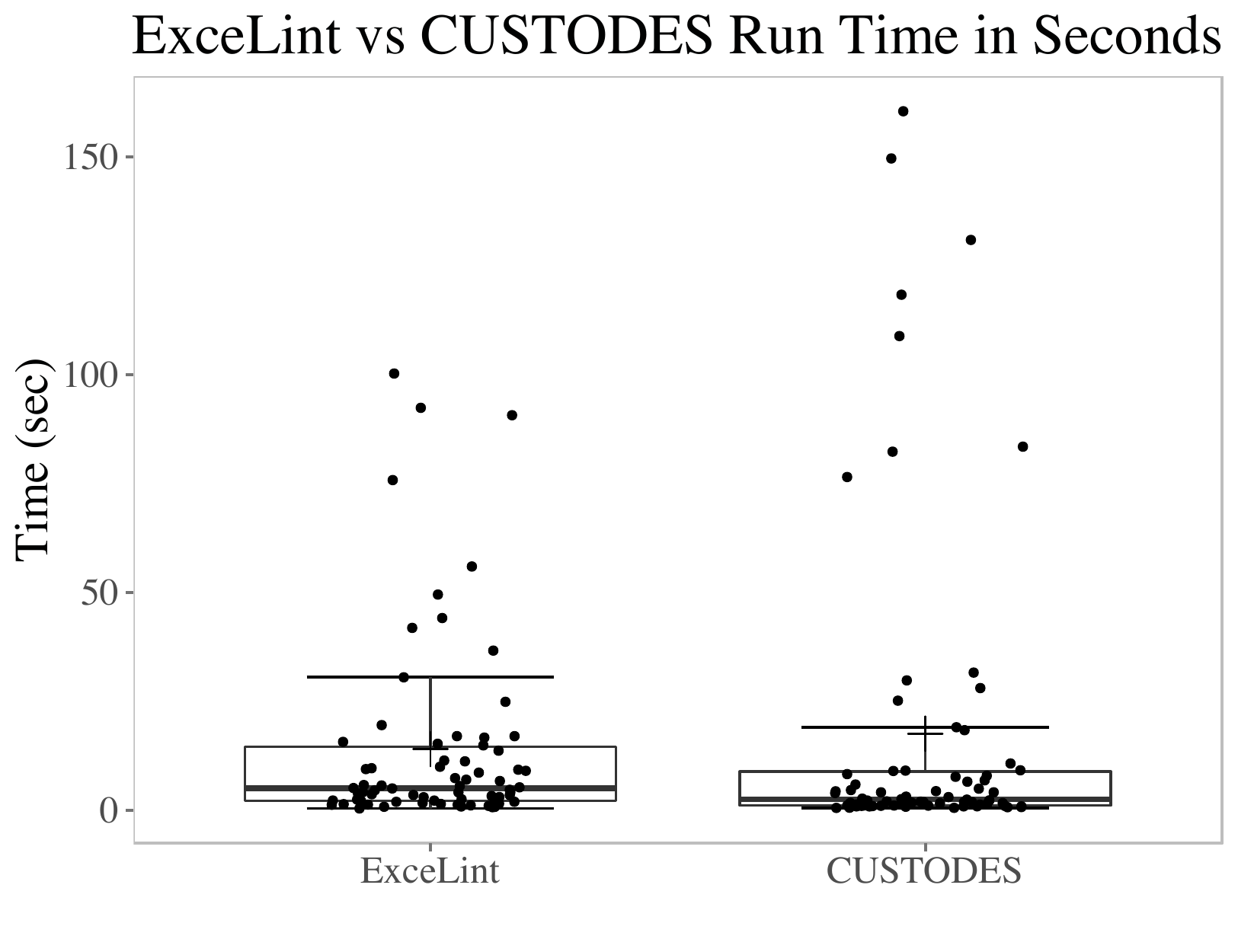}
   \caption{}
   \label{fig:RuntimeComparisonChart}
\end{subfigure}
\caption{(a) \excelint's precision is categorically higher than \custodes; recall is comparable. \excelint's median precision and recall on a workbook are 1.  (b) Performance is similar, typically requiring a few seconds. Detailed results are shown in Figures~\ref{fig:ExPrecisionPlot}, \ref{fig:CPrecisionPlot}, \ref{fig:TPvsFPPlot}, and \ref{fig:DetailedRuntimeComparisonChart}.}
\end{figure}

Our methodology reports a largely distinct set of errors from the \custodes work.  This is in part because we distinguish between suspicious cells and cells that are unambiguously wrong (see ``Formula error'' in \sectref{sec:evaldefs}).  In fact, during our annotation, we also observed a large number (\TotalAnno) of unusual constructions (missing formulas, operations on non-numeric data, and suspicious calculations), many of which are labeled by \custodes as ``smells.''  For example, a sum in a financial spreadsheet with an apparent ``fudge factor'' of \texttt{+1000} appended is highly suspect but not unambiguously wrong without being able to consult with the spreadsheet's original author.  We do not report these as erroneous.

\subsection{RQ1: Are Layouts Rectangular?}
\label{sec:rq0}

Since \excelint relies strongly on the assumption that users produce spreadsheets in a rectangular fashion, it is reasonable to ask whether layouts really are rectangular.  Across our benchmark suite, we found that on average \PctSSMeanRectangular (median: \PctSSMedianRectangular) of all fingerprint regions containing data or formulas (excluding whitespace) in a workbook are rectangular.  When looking specifically at formulas, on average \PctSSFormulaMeanRectangular (median: \PctSSFormulaMedianRectangular) of formula regions in a workbook are also rectangular.  This analysis provides strong evidence that users favor rectangular layouts, especially when formulas are involved.

\subsection{RQ2: Does \excelint Find Formula Errors?}
\label{sec:rq2}

\begin{figure*}[!t]
\centering
\includegraphics[width=\linewidth]{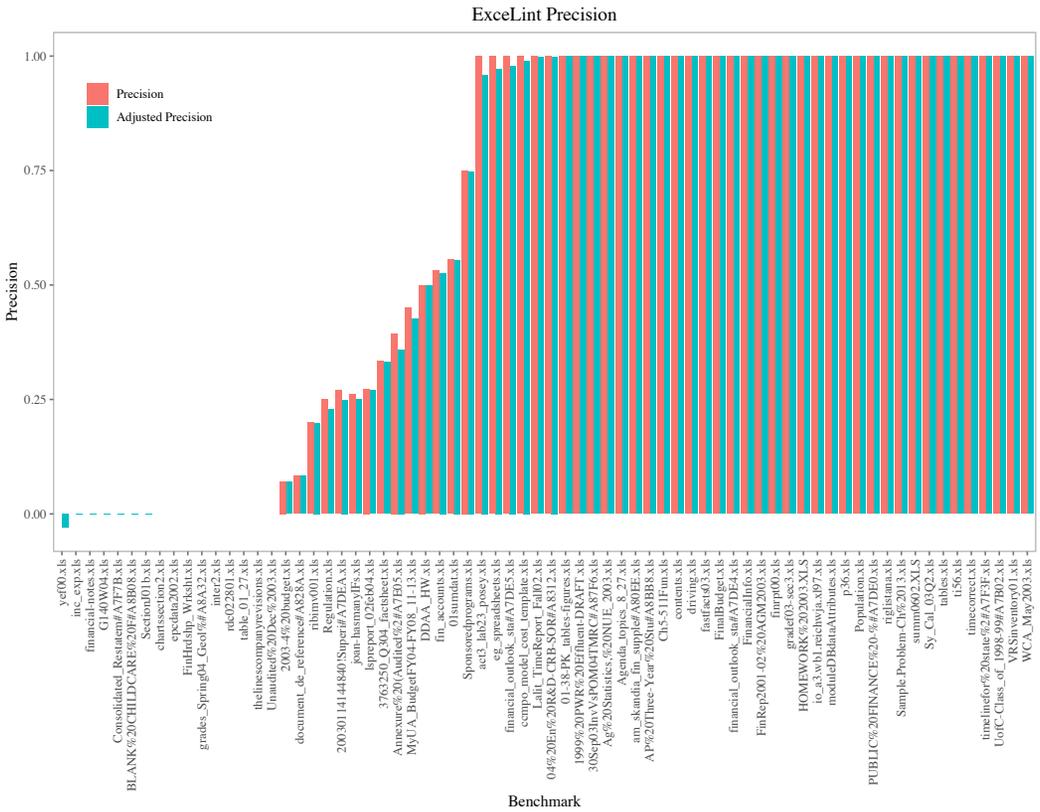}
\caption{\excelint's precision is generally high across the \custodes benchmark suite.  In most cases, adjusting the precision based on the expected number of cells flagged by a random flagger has little effect.  Results are sorted by adjusted precision (see Section~\ref{sec:rq2}).\label{fig:ExPrecisionPlot}}
\end{figure*}

\begin{figure*}[!t]
\centering
\includegraphics[width=\linewidth]{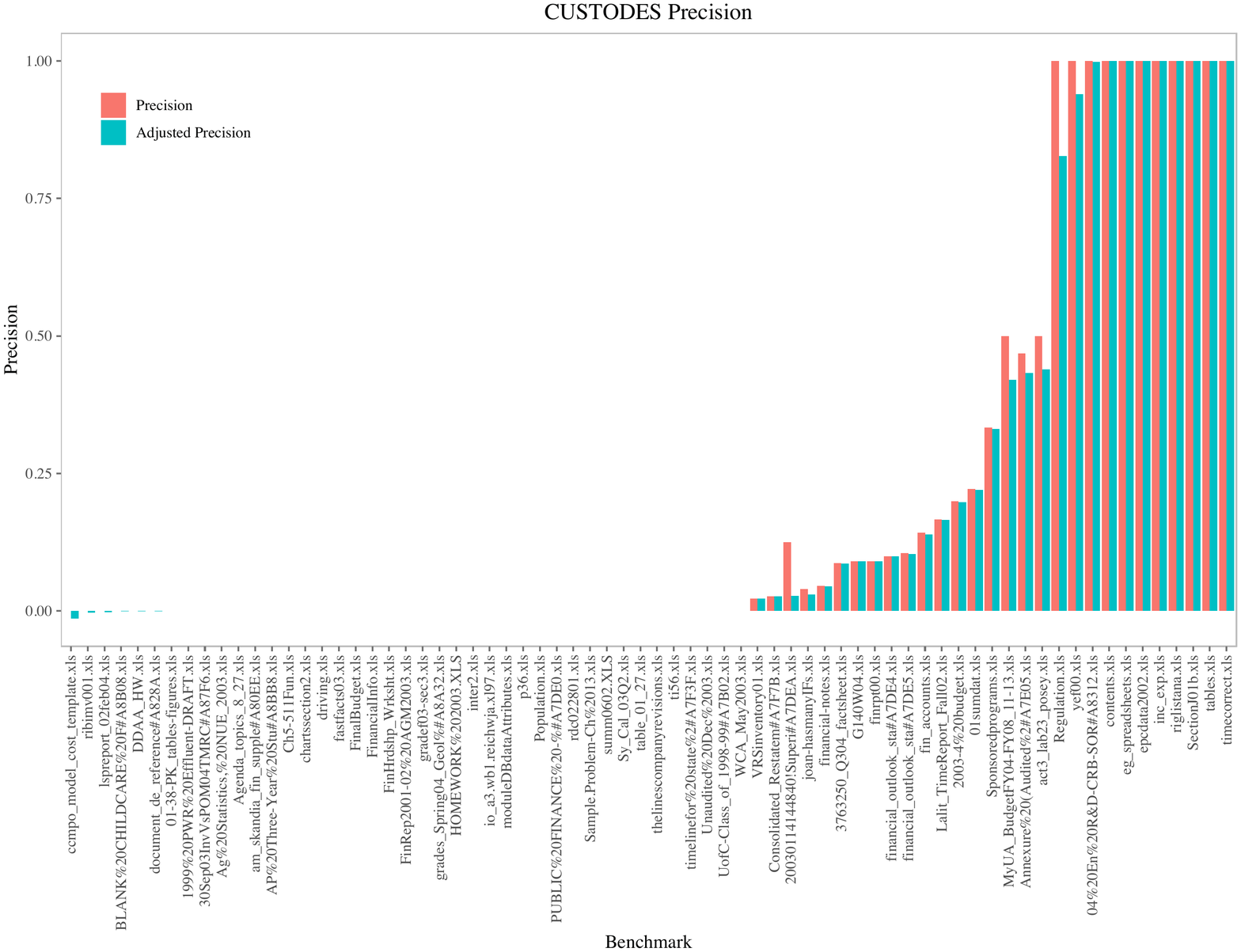}
\caption{\custodes precision is generally lower than \excelint's.  See Figure~\ref{fig:ExPrecisionPlot}.  Results are sorted by adjusted precision (see Section~\ref{sec:rq2}).\label{fig:CPrecisionPlot}}
\end{figure*}

\begin{figure*}[!t]
\centering
\includegraphics[width=\linewidth]{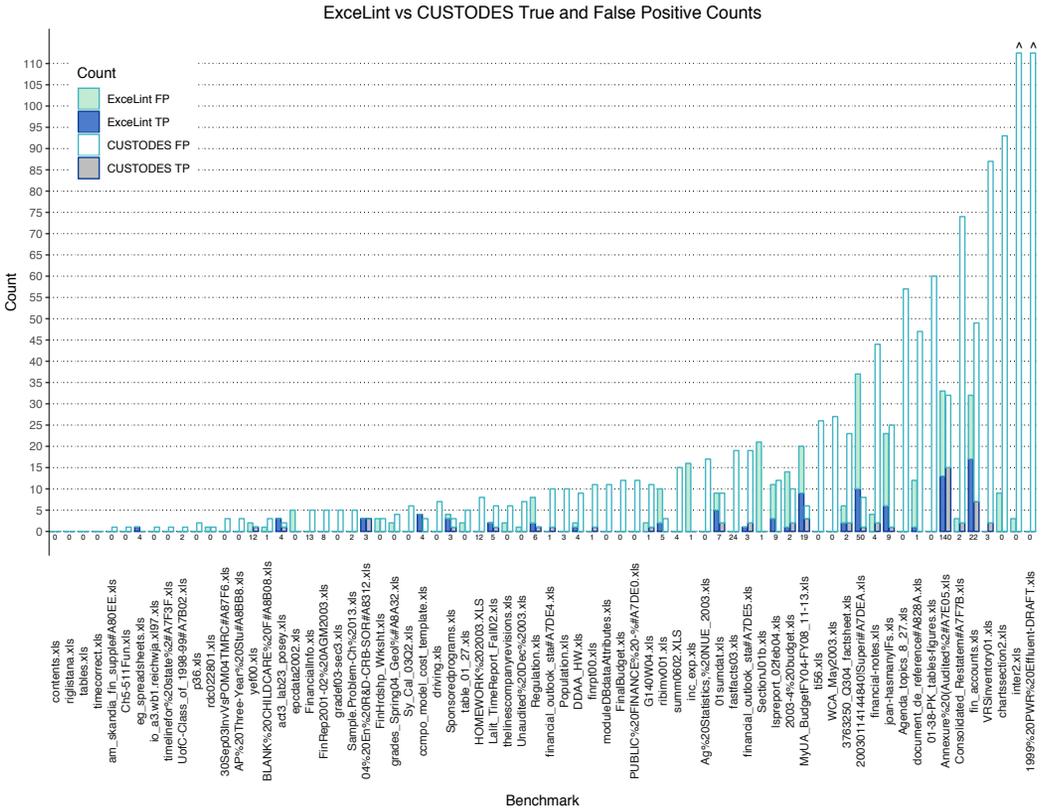}
\caption{\excelint produces far fewer false positives than \custodes.  Critically, it flags nothing when no errors are present.  Each benchmark shows two stacked bars, with \excelint on the left and \custodes on the right.  Numbers below bars denote the ground truth number of errors.  Bars are truncated at two standard deviations; \textasciicircum \;indicates that the bar was truncated (409 and 512 false positives, respectively).\label{fig:TPvsFPPlot}}
\end{figure*}

\begin{figure*}[!t]
\centering
\includegraphics[width=\linewidth]{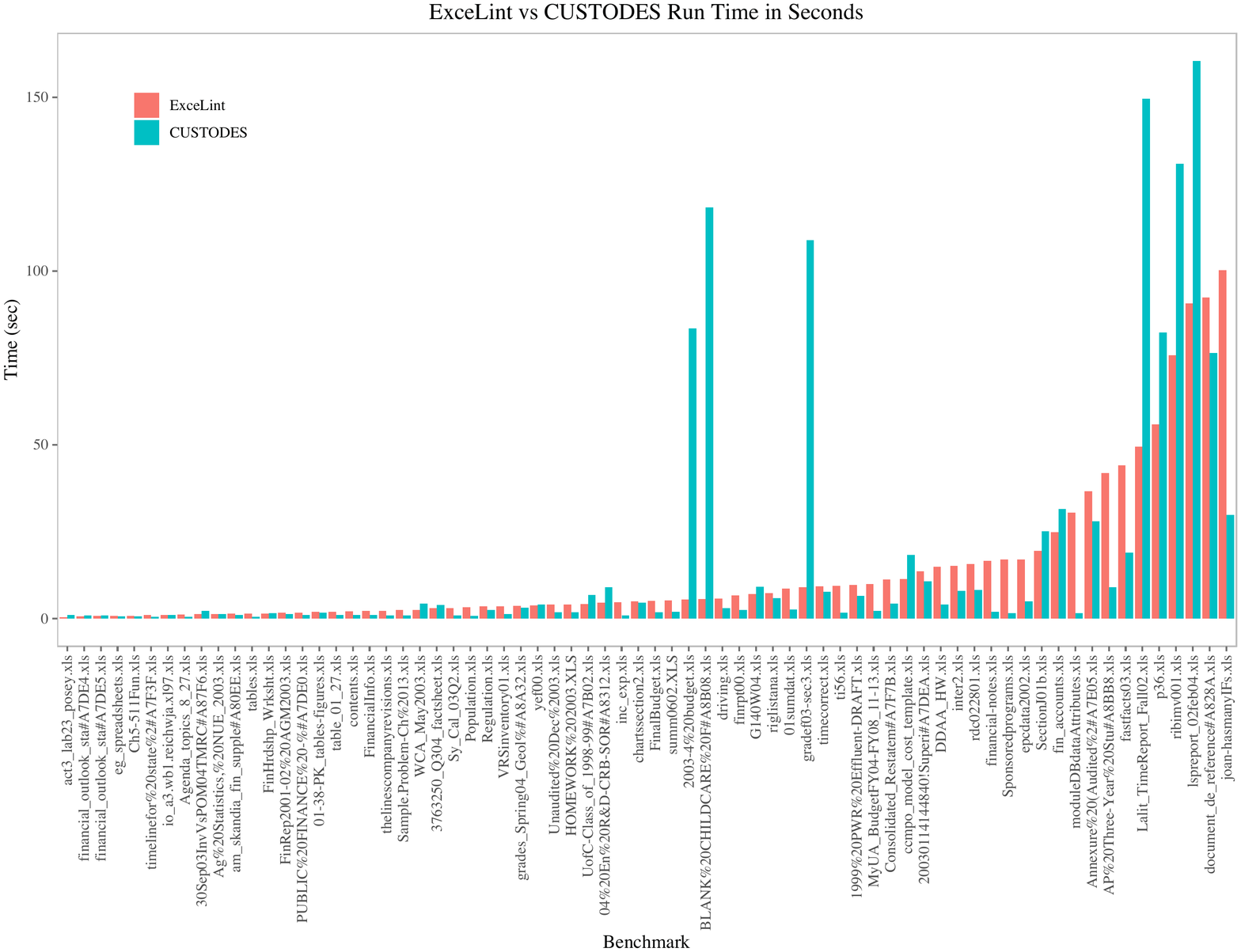}
\caption[\excelint vs \custodes run times.]{\excelint and \custodes have similar performance, typically requiring a few seconds to run an analysis.\label{fig:DetailedRuntimeComparisonChart}}
\end{figure*}

In this section, we evaluate \excelint's guided audit.  Since it is standard practice to compare against the state-of-the-art, we also compare against an alternative tool, \custodes~\cite{custodes}.  \custodes utilizes a pattern-based approach to find ``smelly'' formulas.  While smells are sometimes errors, \custodes is not an error finder.  We argue neither for nor against the merits of finding smells, which could be useful for making spreadsheets more maintainable.  Nonetheless, our goal is to discover \emph{errors}. To our knowledge, \custodes is both the best automated error finder available to the general public. 



\paragraph{Procedure:} A highlighted cell uncovers a real error if either (1) the flagged cell is labeled as an error in our annotated corpus or (2) it is labeled as a bug dual.  We count the number of true positives using the procedure described earlier (see ``Counting errors'').  We use the same procedure when evaluating \custodes.

\paragraph{Definitions:} Precision is defined as $TP / (TP + FP)$ where $TP$ denotes the number of true positives and $FP$ denotes the number of false positives.  When a tool flags nothing, we define precision to be $1$, since the tool makes no mistakes.  When a benchmark contains no errors but the tool flags anything, we define precision to be $0$ since nothing that it flags can be a real error.  Recall is defined as $TP / (TP + FN)$ where $FN$ is the number of false negatives.

\subsubsection*{Difficulty of Accurately Computing Recall} It is conventional to report precision along with recall.  Nonetheless, recall is inherently difficult to measure when using real-world spreadsheets as benchmarks.  To accurately compute recall, the true number of false negatives must be known.  A false negative in our context records when a tool incorrectly flags a cell as not containing an error when it actually does.  The true number of errors in a spreadsheet is difficult to ascertain without a costly audit by domain experts.  We compute recall using the false negative count obtained from our assisted audit; given the large number of suspicious cells we identified, we believe that a domain expert would likely classify more formulas as containing formula errors.  Since we adopt a conservative definition for error, it is likely that the real recall figures are \emph{lower} than what we report.


\paragraph{Results:} Across all workbooks, \excelint has a mean precision of \ExceLintMeanPrecisionPercent (median: \ExceLintMedianPrecisionPercent) and a mean recall of \ExceLintMeanRecallPercent (median: \ExceLintMedianRecallPercent) when finding formula errors.  Note that we strongly favor high precision over high recall, based on the observation that users find low-precision tools to be untrustworthy~\cite{Bessey:2010:FBL:1646353.1646374}.

\begin{sloppypar}
	In general, \excelint outperforms \custodes.  We use \custodes' default configuration. \custodes's mean precision on a workbook is \CUSTODESMeanPrecisionPercent (median: \CUSTODESMedianPrecisionPercent) and mean recall on a workbook is \CUSTODESMeanRecallPercent (median: \CUSTODESMedianRecallPercent). Figure~\ref{fig:wewin} compares precision and recall for \excelint and \custodes.  Note that both tools are strongly affected by a small number of benchmarks that produce a large number of false positives.  For both tools, only \NumHighFPBenchmarks benchmarks account for a large fraction of the total number of false positives, which is why we also report median values which are less affected by outliers. \excelint's five worst benchmarks are responsible for \PctHighFPBenchmarksExcelint of the error; \custodes five worst are responsible for \PctHighFPBenchmarksCUSTODES of the error.  Figure~\ref{fig:TPvsFPPlot} shows raw true and false positive counts.
	
		\custodes explicitly sacrifices precision for recall; we believe that this is the wrong tradeoff.  The total number of false positives produced by each tool is illustrative.  Across the entire suite, \excelint produced \ExceLintTotalTP true positives and \ExceLintTotalFP false positives; \custodes produced \CUSTODESTotalTP true positives and \CUSTODESTotalFP false positives.  Since false positive counts are a proxy for wasted user effort, \custodes wastes roughly \WastedEffort more user effort than \excelint.  
\end{sloppypar}

\paragraph{Random baseline:} Because it could be the case that \excelint's high precision is the result of a large number of errors in a spreadsheet (i.e., flagging nearly anything is likely to produce a true positive, thus errors are easy to find), we also evaluate \excelint and \custodes against a more aggressive baseline.  The baseline is the expected precision obtained by randomly flagging cells.

We compute random baselines analytically.  For small spreadsheets, sampling with replacement may produce a very different distribution than sampling without replacement.  A random flagger samples without replacement; even a bad tool should not flag the same cell twice.  Therefore, we compute the expected value using the hypergeometric distribution, which corrects for small sample size.  Expected value is defined as $\mathop{\mathbb{E}}[X] = n \frac{r}{m}$ where $X$ is a random variable representing the number of true positives, $m$ is the total number of cells, $r$ is the number of true reference errors in the workbook according to the ground truth, and $n$ is the size of the sample, i.e., the number of errors requested by the user.  For each tool, we fix $n$ to be the same number of cells flagged by the tool.

We define $TP_a$, the adjusted number of true positives, to be $TP - \mathop{\mathbb{E}}[X]$.  Correspondingly, we define the adjusted precision to be $1$ when the tool correctly flags nothing (i.e., there are no bugs present), and $\frac{TP_a}{TP_a + FP_a}$ otherwise.

\subsubsection*{Results} \excelint's mean adjusted precision is \ExceLintMeanAdjustedPrecisionPercent and \custodes's mean adjusted precision is \CUSTODESMeanAdjustedPrecisionPercent.  In general, this suggests that neither tool's precision is strongly dictated by a poor selection of benchmarks.  The random flagger performs (marginally) better than \excelint in only \ExceLintRandomNegativeCases of the \numexsheets cases.  The random flagger outperformed \custodes in \CUSTODESRandomNegativeCases cases.

\subsubsection*{High Precision When Doing Nothing} Given that a tool that flags nothing obtains high precision (but very low recall), it is fair to ask whether \excelint boosts its precision by doing nothing.  Indeed, \excelint flags no cells in \ExcelintNumDoNothing of \numexsheets cases.  However, two facts dispel this concern.  First, \excelint has a high formula error recall of \ExceLintMeanRecallPercent.  Second, in \ExcelintNumDoNothingCorrectly cases where \excelint does nothing, the benchmark contains no formula error.  Thus, \excelint saves users auditing effort largely when it should save them effort.


\begin{figure}[!t]
\centering
\includegraphics[width=\linewidth]{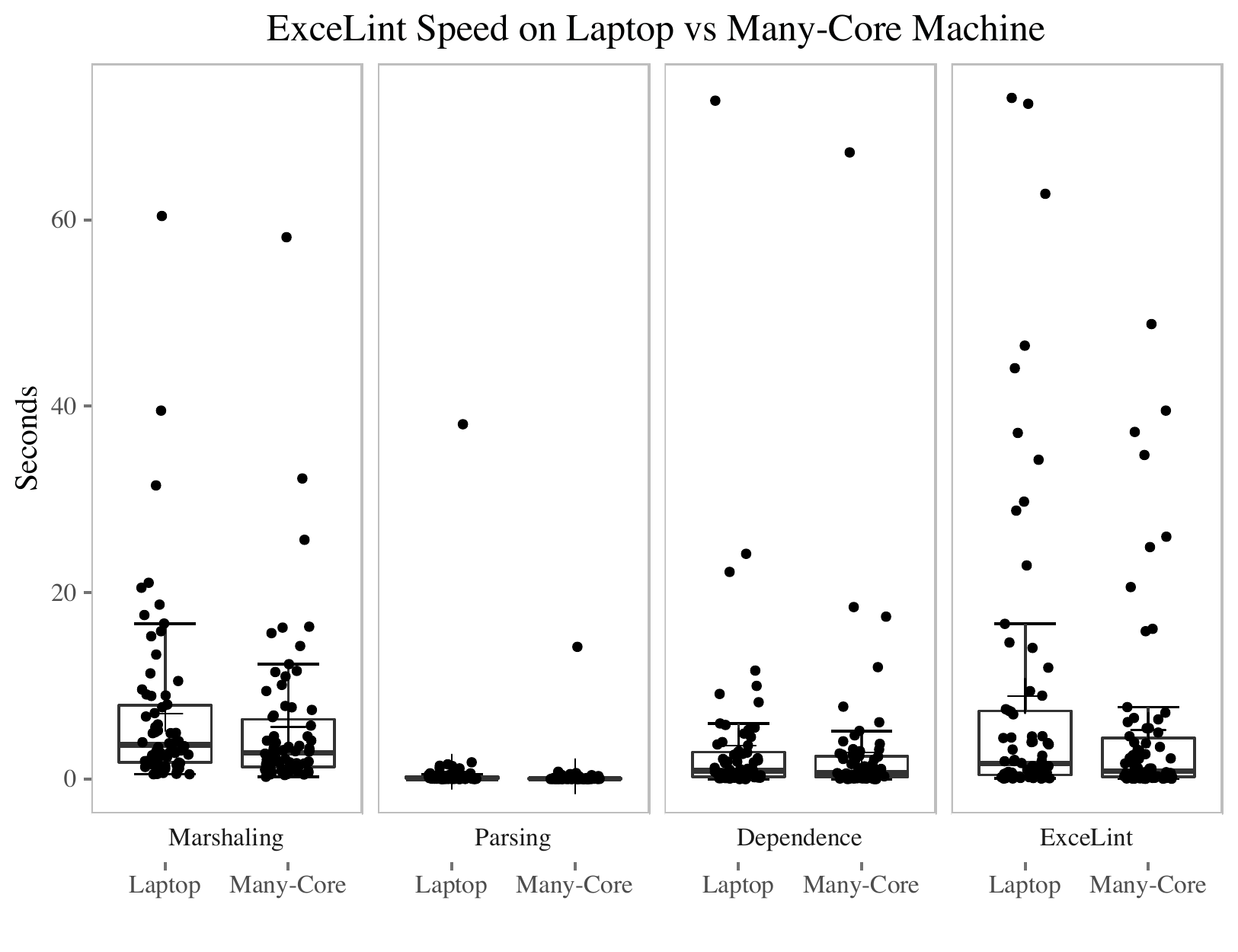}
\caption{Performance measurements for each phase of \excelint's analysis across the benchmark suite using two configurations: a laptop and a multicore machine.  On the multicode machine, \excelint sees the greatest speedup in the entropy decomposition phase of the \excelint analysis, which is multithreaded.\label{fig:RuntimeLaptopVsManycore}}
\end{figure}

\subsubsection*{\custodes' Low Precision} Our results for \custodes are markedly different than those reported by the authors for the same suite (precision: 0.62).  There are a number of reasons for the difference.  First, the goal of this work is to precisely locate unambiguous errors.  \custodes instead looks for suspect patterns.  Unambiguous errors and suspect patterns are quite different; in our experience, suspect patterns rarely correspond to unambiguous errors.  Second, our experience using the \custodes tool leads us to believe that it is primarily tuned to find missing formulas.  While missing formulas (i.e., a hand-calculated constant instead of a formula) are indeed common, we found that they rarely resulted in performing the wrong calculation.  Finally, because of the presence of bug duals, we report fewer bugs than the \custodes group (which reports smells), because our count is intended to capture debugging effort.

\subsection{RQ3: Is \excelint Fast?}
\label{sec:rq3}

\excelint analyzes spreadsheets quickly. \excelint's mean runtime is \ExceLintMeanRuntime seconds, but the mean is strongly affected by the presence of four long running outliers.  \excelint's median runtime across all workbooks is \ExceLintMedianRuntime seconds.  \ExceLintPctUnderThirtySec of the analyses run under 30 seconds. Figure~\ref{fig:RuntimeComparisonChart} compares \excelint's and \custodes' runtimes.

Two factors have a big impact on ExceLint's performance: (1) complex dependence graphs, and (2) dense spreadsheets with little whitespace. The former is a factor because a reference vector must be produced for every reference in a workbook (\sectref{sec:vectorcompute}). The latter is a factor because the entropy calculation cannot be effectively parallelized when the user creates ``dense'' layouts (\sectref{sec:gridpreprocessing}). The longest-running spreadsheets took a long time to produce both reference vectors and to compute entropy, suggesting that they had complex formulas and dense layouts.

	Figure~\ref{fig:RuntimeLaptopVsManycore} compares \excelint's performance on two different hardware configurations: (1) running in VMWare on a typical laptop (a 2016 Macbook Pro) and (2) the multicore desktop machine used for the rest of these benchmarks.  The two configurations exhibit an interesting feature: the median runtime on the laptop is the same as on the high-performance machine.  The reason is that small benchmarks run quickly on both machines, since these benchmarks are dominated by single-thread performance.  Larger benchmarks, on the other hand, benefit from the additional cores available on the high-performance machine.  This difference is most noticeable during the entropy decomposition phase of \excelint's analysis, which is written to take advantage of available hardware parallelism.  As computing hardware is expected to primarily gain cores but not higher clock rates, \excelint's analysis is well-positioned to take advantage of future hardware.

\subsection{RQ4: Case Study: Does \excelint Reproduce Findings of a Professional Audit?}
\label{sec:reinhartrogoff2}

\begin{figure}[!t]
\centering
\includegraphics[width=0.75\linewidth]{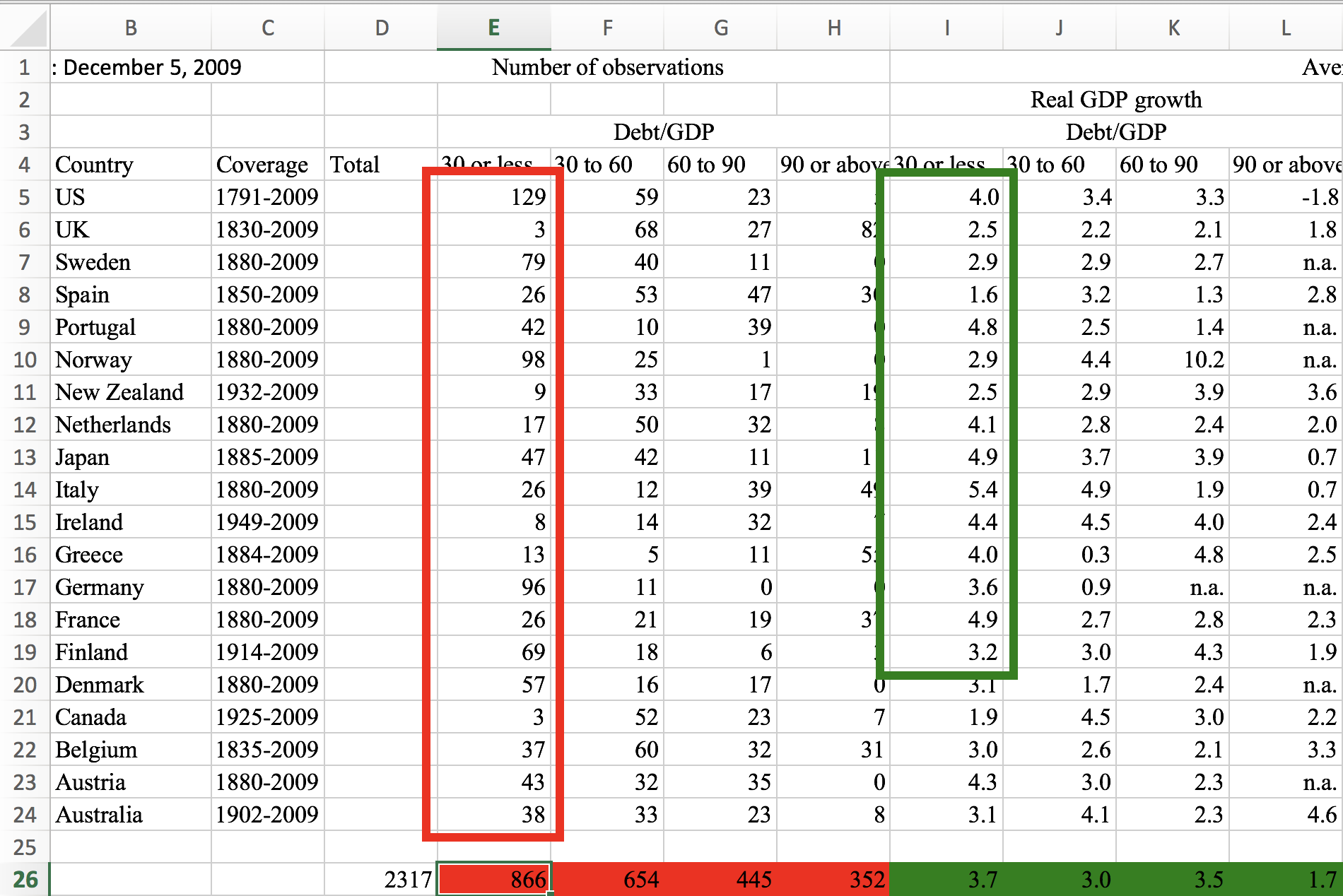}
\caption{\excelint flags the formulas in \texttt{E26:H26}, which are inconsistent with formulas in \texttt{I26:X26}.  The red and green boxes show the set of cells referenced by \texttt{E26} and  \texttt{I26}, respectively.  While \excelint marks \texttt{E26:H26} as the ``error'' because it is the smaller set, in fact, \texttt{E26:H26} are correct.  This spreadsheet contains a systematic error and all formulas in \texttt{I26:X26} are incorrect.  \label{fig:RRAnomaly}}
\end{figure}

In 2010, the economists Carmen Reinhart and Kenneth Rogoff published the paper ``Growth in a Time of Debt'' (GTD) that argued that after exceeding a critical debt-to-GDP ratio, countries are doomed to an economic ``death spiral''.  The paper was highly influential among conservatives seeking to justify austerity measures.  However, it was later discovered by Herndon et al. that GTD's deep flaws meant that the opposite conclusion was true.  Notably, Reinhart and Rogoff utilized a spreadsheet to perform their analysis.

Herndon et al. call out one class of spreadsheet error as particularly significant~\cite{herndon2013does}.  In essence, the computation completely excludes five countries---Denmark, Canada, Belgium, Austria, and Australia---from the analysis.

We ran \excelint on the Reinhart-Rogoff spreadsheet and it found this error, repeated twice.  Both error reports were also found by professional auditors.  Figure~\ref{fig:RRAnomaly} shows one of the sets of errors.  The cells in red, \texttt{E26:H26}, are inconsistent with the set of cells in green, \texttt{I26:X26}.  In fact, \texttt{I26:X26} is wrong and \texttt{E26:H26} is correct, because \texttt{I26:X26} fails to refer to the entire set of figures for each country, the cells in rows 5 through 24.  Nonetheless, by highlighting both regions, it is immediately clear which of the two sets of cells is wrong.

\subsection{Summary}

Using the \excelint global view visualization, we uncovered \ECTrueRefBugDelta more errors than \custodes, for a total of \ExceLintNumTrueRefBugs errors, when used on an existing pre-audited corpus.  When using \excelint to propose fixes, \excelint is \ExceLintPrecisionPointDelta percentage points more precise than the comparable state-of-the-art smell-based tool, \custodes.  Finally, \excelint is fast enough to run interactively, requiring a median of \ExMedianRuntimeSeconds seconds to run a complete analysis on an entire workbook.

\section{Related Work}
\label{sec:related}
\paragraph{Smells:}
One technique for spreadsheet error detection employs \emph{ad hoc} pattern-based approaches. These approaches, sometimes called ``smells'', include patterns like ``long formulas'' that are thought to reflect bad practices~\cite{hermans2012detecting}. Excel itself includes a small set of such patterns. Much like source code ``linters,'' flagged items are not necessarily errors.  While we compare directly against only one such tool---\custodes---other recent work from the software engineering community adopts similar approaches~\cite{custodes,Hermans2015,dou2014spreadsheet,DBLP:journals/tr/HoferHW17,DBLP:journals/jss/JannachSHW14}. For example, we found another smell-based tool, \faultysheet, to be unusably imprecise~\cite{6976077}.  When run on the ``standard solution'' provided with its own benchmark corpus---an error-free spreadsheet---\faultysheet flags 15 of the 19 formulas present, a 100\% false positive rate.  Both \excelint and \custodes correctly flag nothing.

\paragraph{Type and Unit Checking:} Other work on detecting errors in spreadsheets has focused on inferring units and relationships (\emph{has-a}, \emph{is-a}) from information like structural clues and column headers, and then checking for inconsistencies~\cite{Antoniu:2004:VUC:998675.999448,erwig2002adding,DBLP:conf/kbse/AhmadAGK03,Chambers:2010:RSL:1860134.1860346,Erwig:2009:SES:1608570.1608694,Erwig:2005:AGM:1062455.1062494,StateOfTheArt,abraham2004header}.  These analyses do find real bugs in spreadsheets, but they are largely orthogonal to our approach.  Many of the bugs that \excelint finds would be considered type- and unit-safe.

\paragraph{Fault Localization and Testing:} Relying on an error oracle, \emph{fault localization} (also known as \emph{root cause analysis}) traces a problematic execution back to its origin.  In the context of spreadsheets, this means finding the source of an error given a manually-produced annotation, test, or specification~\cite{Hofer:2013:EEF:2450312.2450321,DBLP:journals/ase/HoferPAW15,DBLP:journals/sqj/AbreuHPW15}.  Note that many localization approaches for spreadsheets are evaluated using randomly-generated errors, which are now known not generalize to real errors~\cite{Pearson:2017:EIF:3097368.3097441}.  Localization aside, there has also been considerable work on testing tools for spreadsheets~\cite{fisher2006scaling,rothermel1998you,rothermel2001methodology,Carver:2006:EET:1159733.1159775,StateOfTheArt,DBLP:conf/vl/SchmitzJHKSW17,DBLP:conf/pts/AbreuAHW15}

\excelint is a fully-automated error finder for spreadsheets and needs no specifications, annotations, or tests of any kind.  \excelint is also evaluated using real-world spreadsheets, not randomly-generated errors.

\paragraph{Anomaly Detection:} An alternative approach frames errors in terms of anomalies.  Anomaly analysis leverages the observation from conventional programming languages that anomalous code is often wrong~\cite{Engler:2001:BDB:502034.502041,Xie02usingredundancies, Hangal:2002:TDS:581339.581377,Chilimbi:2006:HIH:1168857.1168885, Raz:2002:SAD:581339.581378,Dimitrov:2009:ABP:1508244.1508252}.  This lets an analysis circumvent the difficulty of obtaining program correctness rules.

\excelint bears a superficial resemblance to anomaly analysis.  Both are concerned with finding unusual program fragments.  However, \excelint's approach is not purely statistical; rather, the likelihood of errors is based on the effect a fix has on the entropy of the layout of references.

\section{Conclusion}
\label{sec:conclusion}
This paper presents \excelint, an information-theoretic static
analysis that finds formula errors in spreadsheets.  We show
that \excelint has high precision and recall (median: 100\%). We have
released \excelint as an open source project~\cite{excelint-github}
that operates as a plugin for Microsoft Excel.

\section*{Acknowledgments}
This material is based upon work supported by the National Science Foundation under Grant No. CCF-1617892.

\afterpage{\FloatBarrier}
\balance
\bibliography{refs}


\end{document}